\documentclass{aa}  
\usepackage{graphicx}
\usepackage{lscape}
\usepackage{hyperref}
\usepackage{txfonts}
\usepackage{float}

\newcommand{\injindx}{\alpha_{\text{inj}}}

\begin{document}
	\title{The ultra low-frequency spectral properties of bright extended radio galaxies in the 3CRR catalogue}

   \author{J. M. Boxelaar\inst{1}\fnmsep\inst{2}\fnmsep\thanks{\email{j.boxelaar@ira.inaf.it}}
          \and
          F. De Gasperin\inst{1}
          \and
          M. J. Hardcastle\inst{3}
          \and
          E. De Rubeis\inst{4}\fnmsep\inst{1}
          \and
          R. J. van Weeren\inst{5}
          }

   \institute{INAF–Istituto di Radioastronomia, Via P. Gobetti 101, 40129 Bologna, Italy
         \and
             Dipartimento di Fisica e Astronomia, Università di Bologna, via P. Gobetti 93/2, 40129 Bologna, Italy
        \and
            Centre for Astrophysics Research, University of Hertfordshire, College Lane, Hatfield AL10 9AB, UK
        \and
            Hamburger Sternwarte, Universit\"at Hamburg, Gojenbergsweg 112, 21029 Hamburg, Germany
        \and
            Leiden Observatory, Leiden University, PO Box 9513, 2300 RA Leiden, The Netherlands
             }

   \date{Received \today; accepted NA}

\abstract{Active galactic nuclei (AGN) jets are fundamental drivers of galaxy evolution, injecting kinetic energy into their environments. The large-scale morphology and spectral properties of these radio galaxies are consequences of complex particle acceleration, energy loss, and absorption processes. While the shape of the synchrotron spectrum encodes the plasma's energetic history, understanding the physics of particle acceleration and duty cycles has historically been limited by a lack of well-resolved observations at ultra-low frequencies ($<100$~MHz), where the oldest cosmic ray electron populations are traced.}
{This study aims to perform the first comprehensive multi-frequency analysis of bright extended radio galaxies down to 58~MHz. The goal is to study electron acceleration mechanisms, accurately measure the low-frequency spectral shape, and constrain the injection index ($\injindx$) for a sample of Fanaroff-Riley (FR) I and II galaxies using spectral ageing models.}
{Utilising new 58~MHz observations from the LOFAR Low Band Antenna (LBA) combined with LOFAR High Band Antenna (HBA; 144 MHz) and Rapid ASKAP Continuum Survey (RACS, 887~MHz \& 943.5~MHz \& 1367.5~MHz) data, a sub-sample of 22 extended sources from the 3CRR catalogue was selected, requiring the largest angular size to be at least 2.5\arcmin. The analysis involves constructing detailed spectral index maps and utilising radio colour-colour diagrams to interpret spectral shapes and constrain ageing model parameters across the radio lobes.}
{This study presents the ultra-low frequency spectral index maps for this sample. For FR I galaxies, spectral indices range from $\sim0.5$ near the core (consistent with first-order Fermi acceleration) to $>1.0$ in the lobes. For FR II galaxies, hotspots exhibit steep low-frequency spectra ($0.5-0.9$), suggesting complex acceleration or absorption effects.
}
{The addition of ultra-low frequency data is critical for understanding radio galaxy age, energetics, and accurately constraining the injection index. While some sources follow standard ageing models in colour-colour space, those with complex morphologies often deviate from them, indicating that simple ageing models may not be applicable to all radio galaxy structures.}

   \keywords{Radio continuum: galaxies -- Techniques: interferometric -- Galaxies: active -- Galaxies: jets -- Catalogs -- Surveys}

   \maketitle

\section{Introduction}
\label{sec:introduction}
Active galactic nuclei (AGN) jets play a central role in regulating galaxy formation and evolution through the injection of kinetic energy into their surrounding environments, and they are are believed to be among the most influential drivers of galaxy evolution \citep{2006MNRAS.365...11C, 2006MNRAS.370..645B, 2009MNRAS.395..518C}. Besides this, the formation of hundreds of kilo- to megaparsec scale lobes also injects kinetic energy into much larger scales such as those of groups and clusters of galaxies \citep{2008MNRAS.383..525K,2021NatAs...5.1261B}. However, the mechanisms by which jets accelerate these particles to relativistic energies, how efficiently it couples to their environments, and how they evolve over their lifetimes are poorly understood. 

The large-scale morphology of AGN at radio frequencies is a result of the complex physics of particle acceleration and matter accretion by the central black hole and the interaction between the ejected plasma lobes and the environment. Studying these extended structures provides fundamental insight into the underlying physical processes. The shape of the synchrotron spectrum contains information on both acceleration and loss processes, while the variations in the localised spectra across the lobes contain information on the characteristic age and life cycle of the radio galaxies \citep[e.g.][]{Brienza2020}. 

Generally, two morphologically distinct types of radio galaxies are identified: Fanaroff–Riley (FR) class I and II galaxies \citep{Fanaroff1974}. The former refers to centre brightened structures where the radio lobes decrease in brightness further away from the central AGN, which is believed to be due to deceleration and de-collimation of the jet \citep{1994ApJ...422..542B, 2014MNRAS.437.3405L}. The latter of the two classes, on the other hand, are edge brightened, and almost always with distinctive hotspots at the terminal point of the jets. Another physical distinction between the two FR types is that for FR~Is, the acceleration happens close to the central AGN while for FR~IIs the acceleration is also manifested at the hotspots \citep[e.g.][]{Alexander1987MNRAS.225...27A,Katz-Stone1993}. 

Despite their importance, our understanding of the physics, energetics, and duty cycles of radio galaxies has historically been based on small samples of bright, nearby objects due to the difficulty to constrain the key dynamic properties \citep[e.g.][]{2008MNRAS.383..525K, 2008MNRAS.385.1286J}. This is largely due to the fact that the radiating plasma records a complex history of particle acceleration, radiative ageing, and magnetic-field evolution that is not directly observable at a single frequency or epoch. Wide-area radio surveys are now transforming this picture. Large area surveys such as the Low Frequency Array \citep[LOFAR;][]{vanHaarlem2013} Two Metre Sky Survey \citep[LoTSS;][]{Shimwell2019}, the MeerKAT MIGHTEE survey \citep{2012AfrSk..16...44J}, and the Rapid ASKAP Continuum Survey \citep[RACS,][]{RACS2020} provide large, homogeneous samples spanning a wide dynamic range in luminosity, redshift, and environment. However, a comprehensive study of resolved radio galaxies below 100~MHz is missing. This study presents the first such comprehensive multi-frequency analysis of radio galaxies down to 58~MHz using the LOFAR telescope. 

The Low Band Antenna (LBA) system of LOFAR, operating between 10 and 90~MHz, in particular provides access to a spectral regime that has remained largely unexplored at angular resolutions below 30\arcsec. These frequencies trace the oldest electron populations — particles whose radiative lifetimes may exceed tens of millions of years — and therefore encode crucial information about past jet activity, duty cycles, and the integrated energy released by the AGN.  
A recent work by \citet{Boxelaarprep} produced maps of the entire Third Cambridge Catalogue of Radio Sources \citep[3CRR;][]{Laing1983} of the brightest radio galaxies in the northern sky. The 3CRR catalogue consists of the brightest extragalactic radio sources with flux densities $S_{\nu}> 10$~Jy at 178~MHz. Sources are constrained to have a declination $\delta>10\degr$, and to be more than $10\degr$ below or above the galactic plane. This sample contains 173 sources, among which are the most extended and bright FR~I and IIs in the sky, and it provides an excellent opportunity to utilise the lowest frequencies to study the spectral and dynamical properties of bright radio galaxies. 

For many years, workers in the field have tried to characterise the acceleration mechanism in radio galaxies by measuring the injection index $\injindx$ \citep{Katz-Stone1993, 1997ApJ...488..146K,1999A&A...349..381H,2005ApJ...626..748Y,2013MNRAS.432.1114L}, which as the name suggests is directly related to the energy spectrum of freshly injected plasma into the medium either at the core (for FR~I) or at the hotspots (for FR~II). The initial energy number density of the electrons injected into the medium is $N(E) =N_0E^{-\delta}$ \citep{Pacholczyk1970}, where $\delta$ is related to the injection index through $\delta = 2\alpha_{inj}+1$. Different models of particle acceleration predict different distributions of values of $\injindx$ that can be measured. First-order Fermi acceleration in strong, non-relativistic shocks \citep{1978MNRAS.182..147B}, for example, predict $\injindx=0.5$, while the limit for ultra-relativistic shocks is 0.615 \citep{2000ApJ...542..235K} and mildly relativistic shocks can produce a variety of spectral slopes (see \citet{2012ApJ...745...63S, 2013MNRAS.432.1114L} for a more comprehensive discussion). These works often expressed the need for more high-resolution low-frequency observations to better constrain the injection indices and be more conclusive regarding the possible acceleration mechanisms. In a previous work, we produced images at 15\arcsec\ resolution at a frequency of 58~MHz for the entire 3CRR sample \citep[][ hereafter Paper~I]{Boxelaarprep}. This allowed us to accurately measure $\injindx$ with high-resolution radio data at frequencies $<100$~MHz for the first time.  

This paper presents the methods, prospects and limitations for studying electron acceleration mechanisms with future LBA observations. We took a sub-sample containing the most extended sources presented in Paper~I and combine it with higher frequency data from LOFAR HBA at 144~MHz and ASKAP-low and ASKAP-mid to comprehensively measure the low frequency spectral index and constrain the injection index based on ageing models (see the next section for the definition). The sub-sample contains 22 3CRR sources and this allows us to present the most comprehensive comparison yet of the ultra-low frequency spectral structure of bright FR~I and FR~II galaxies. In Section~\ref{sec:observation} we further describe the sample and source selection, Section~\ref{sec:spidx} shows the resulting spectral index maps and injection index measurements. We subsequently discuss the results and caveats of the analysis in Section~\ref{sec:discussion} and list the conclusions in Section~\ref{sec:conclusion}.

\label{sec:ageing}
A key aspect of the study of radio galaxies is understanding their dynamical history and evolution. Through this evolution, we try to understand the (re-)acceleration mechanisms that allows the observation of up to mega parsec scale radio lobes \citep{2024A&A...686A..21S,2025A&A...699A.257A}. Spectral analysis of radio galaxies reveals their dynamical history through observed synchrotron energy losses \citep[see e.g.][]{2008MNRAS.385.1286J, Orru2010, 2017MNRAS.469..639H,2018MNRAS.474.3361T,2024AdSpR..73.1113P}. Studying AGN plasma properties through the spectral signatures of synchrotron energy losses, also known as synchrotron ageing, has become common practice and allows us to infer the approximate age of the plasma as well as deriving order of magnitude estimations of the magnetic field strength \citep[e.g.][]{Harwood2013MNRAS.435.3353H, Hardcastle2013MNRAS.433.3364H, Harwood2015MNRAS.454.3403H}. Low frequency observations are especially useful in spectral modelling as the slope of the low frequency end of the spectrum constrains one of the model parameters: the injection index $\alpha_{inj}$.

Generally, five mechanisms of energy loss are considered: ionisation of the local medium, free-free radiation, synchrotron radiation, inverse Compton radiation \citep{Pacholczyk1970} and finally adiabatic losses due to expansion or compression, which become relevant at kiloparsec scales. The latter two losses have the same energy dependency and are often considered as the only dominant loss of electron energy for relativistic electrons \citep{Harwood2013MNRAS.435.3353H}. The synchrotron radiation loss over time scales as $-A E^2$, where $A$ is a time independent term depending on primarily the magnetic field and the angle of the magnetic field vector and the direction of motion of the electron \citep{Rybicki1979rpa..book.....R}. At higher electron energies (hence higher frequencies) the loss is greater and this is why a steepening towards higher frequency is expected. The two most widely used ageing models are the JP \citep{Jaffe1973} and KP \citep{Kardashev1962,Pacholczyk1970} models, both of which assume a single cosmic ray electron injection at $t=0$. These models are especially useful for the study of resolved lobes and hotspots as there the assumption of a single energy injection has been shown to fit fairly well. This of course relies on the availability of spatially resolved, high fidelity radio observations at multiple frequencies, something that is often unavailable, especially at lower frequencies. With the data presented in this work, it becomes possible for the first time to extend the ageing analysis to the ultra-low frequency regime and study whether these ageing models still hold. 

Because only a small portion of the sample is resolved at the level that the source is spanned by multiple beam sizes, one could consider adopting an alternative to the JP and KP models that is supposedly suitable for the study of unresolved sources. One variation assumes a continuous injection (CI) of new electrons which allows us to model the spectral ageing using the integrated spectrum of the entire radio source \citep{Pacholczyk1970}. The CI model does however have some significant limitations and it has been argued by multiple authors that considering integrated spectra does not reflect the physical properties of the radio source \citep[e.g.][]{2002NewAR..46..105Y, 2005ApJ...626..748Y,2017MNRAS.466.2888H, 2025PASA...42..153J}. Another variation by \citet{1994A&A...285...27K} modifies JP and KP models to take into account the finite lifetime of the injection of relativistic particles. This variation works especially well for relic radio galaxies.

Assuming that a single population of plasma electrons is injected into the lobes with a single magnetic field strength $B$, with an initial power-law number density, the flux density at a fixed frequency $S_{\nu}$ is given by 
\begin{equation}
\label{eq:int}
	S_{\nu} = C_1\int^{\gamma_{\text{max}}}_{\gamma_{\text{max}}}\int^{\pi}_0 F\left(\frac{\nu}{\nu_c}\right)n_e(\gamma)B\sin^2{\theta}\;\mathrm{d}\gamma\mathrm{d}\theta.
\end{equation}
In this equation, $C_1 = \sqrt{3}e^3B/8\pi\epsilon_0cm_e$ \citep[see ][]{Harwood2013MNRAS.435.3353H}, $F(\nu)$ describes the synchrotron emissivity of a single cosmic ray electron in a magnetic field \citep[see][]{Rybicki1979rpa..book.....R}, $n_e$ is the free electron number density as a function of Lorentz factor and $\theta$ is the pitch angle between the magnetic field vector and the velocity vector of the electrons. Following the derivations of \cite{Harwood2013MNRAS.435.3353H, Harwood2015MNRAS.454.3403H} we adopted an electron energy density (as a function of the Lorentz factor $\gamma$) for the JP and KP models given by 
\begin{align}
	\label{eq:JP_ny}
	n(\gamma) = n_0\gamma^{-\delta}(1-A_{\text{JP}}\gamma t)^{-\delta-2},
\end{align}
where $A_{\text{JP}}=\frac{4\sigma_t}{6m_e^2\nu_c^3\mu_0}B^2$ and $\nu_c=3\gamma^2eB_{\perp}/4\pi m_ec$ is the 'critical frequency'. Although the models are slightly different, the resulting injection index is physically the same. For the KP model, $A_{\text{KP}}=A_{\text{JP}}\sin{\theta}$.

Currently accepted ageing models make assumptions on constant magnetic fields throughout the lobe or large portions of it and assume a single injection index, implicitly assuming a constant jet power over the AGN accretion history and a single dominant particle acceleration mechanism. Trying to fit the spectral shape in different parts of a lobe to different ageing models has become common practice. There are a few limitations to this practice \citep[e.g.][]{Harwood2013MNRAS.435.3353H,2018MNRAS.474.3361T, 2018MNRAS.473.4179T,2020MNRAS.491.5015M,2024A&A...691A..76W}. First, it is often difficult to obtain high-fidelity and high-resolution images of sources spanning a decent frequency range. This will naturally limit the number of measurements in frequency space and thus the accuracy of fits with a high degree of freedom for model parameters. Second, low frequency surveys like the VLA Low-Frequency Sky Survey \citep[VLSS, ][]{VLSS2007} at 74~MHz have low resolution ($1.4\arcmin$) and high-res olution surveys exist only above a few hundreds of GHz, often forcing the analysis to high frequency ranges which do not allow for a good constraint on the injection index of the observed plasma.

\section{Observations and sample}
\label{sec:observation}
In this paper we continued the work published recently by \citet{Boxelaarprep}, where we presented the entire 3CRR catalogue observed with LOFAR at 58~MHz. The full description of the 3CRR sample and a comprehensive description of the observations and resulting maps are discussed there \footnote{The LBA and HBA radio maps for all sources can be found at \url{https://www.lofar-surveys.org/3CRR-LBA.html}}. This work takes a subset of that catalogue, complemented by observations at higher frequencies, to study the detailed spectral properties of these radio galaxies.

\begin{table}[ht]
	\caption{\label{tab-all-obs-new} Summary of observations.}
	\centering
	\begin{tabular}{lcccc}
		\hline\hline
		Instrument & Freq.  & Median RMS& Flux scale err. \\
		& [MHz]  & mJy~beam$^{-1}$ & \\
		\hline
		LOFAR LBA\tablefootmark{a} & 57.7&10 & 10\%\\
		LOFAR HBA\tablefootmark{b} & 144 & 0.132 & 10\%\\
		RACS (low)\tablefootmark{c} & 887.5&0.266&7\%\\
		& 943.5&0.205&7\%\tablefootmark{c}\\
		RACS (mid)\tablefootmark{c} & 1367.5&0.198&7\%\\
		RACS (high)\tablefootmark{c} & 1655.5&0.209&7\%\\
		\hline
	\end{tabular}
    \tablefoot{For every observation we note the central observing frequency, the median RMS noise and the flux scale error taken from the relevant reference.}
    \tablebib{(a)~\citet{Boxelaarprep}; (b)~\citet{Shimwell2019}; (c)~\citet{RACS2020}}
\end{table}

\subsection{Sample of extended radio galaxies}
With the new observations at 58~MHz, we have the unique opportunity to study FR~I and FR~II galaxies down to very low frequencies. Since this study requires sources to be rather extended, we have constructed a subsample of 3CRR containing only extended sources, i.e. large sources with a linear angular size (LAS) of at least ten beam-sizes in the LBA data. Furthermore, we required all sources to be covered by the same collection of surveys. The homogeneity requirement is not easy to uphold as there are virtually no surveys besides LoTSS \citep{Shimwell2019} at 144~MHz with a full northern sky coverage, a resolution $\sim20\arcsec$ and sensitivity to arc-minute scale diffuse emission. The survey that comes closest to the requirements is the Rapid ASKAP Continuum Survey (RACS, \citet{RACS2020}), having coverage $<40\degr$~DEC and three frequency bands being RACS-low (887~MHz or 943.5~MHz, \citet{RACS2020}), RACS-mid (1367~MHz, \citet{RACS-mid2023}) and RACS-high (1655~MHz, \citet{RACS-high2025}). Sources from ASKAP-low are observed at either 887~MHz or 943~MHz. We adopt 887~MHz as the representative frequency for RACS-low throughout to avoid confusion. Figures will always indicate the right frequency of the observation of that source. RACS images are reported to have a median noise of 330~$\mu$Jy~beam$^{-1}$. This is summarised in Table~\ref{tab-all-obs-new}. The subsample of sources for which we can carry out the spectral analysis is constrained by: 

\begin{itemize}
	\item Part of the 3CRR sample
	\item Linear size larger than 10 LBA beams (15\arcsec)
	\item Covered by LoTSS DR3
	\item $\text{DEC}<40\degr$ (covered by RACS)
\end{itemize}

\begin{table*}[ht]
	\caption{\label{tab:the_subsample} Subsample of extended sources used in spectral analysis.}
	\centering
	\begin{tabular}{lccccccc}
		\hline\hline
		Object & LAS\tablefootmark{a} & FR type\tablefootmark{a} &$S_{58\text{ MHz}}$ & dyn. range& ${\sigma_{\text{rms}}}$ & Final beam\\
		&  & & Jy &  & mJy~beam$^{-1}$ & $\arcsec\times\arcsec$ \\
		{\bf ASKAP Available} &&&&&&\\
		\hline
3C 192 & 3.34\arcmin & II & 58.7 & 98 & 7.3& $25.0\times25.0$\\
3C 223 & 5.11\arcmin & II &  38.0 & 76 & 9.5&$39.7\times39.7$\\
3C 236 & 40.76\arcmin & II & 38.0 & 95 & 11.7&$41.2\times41.2$\\
3C 264 & 6.60\arcmin & I & 63.9 & 31 & 12.1&$25.0\times25.0$\\
3C 284 & 2.96\arcmin & II & 39.4 & 69 & 8.8& $25.0\times25.0$\\
3C 296 & 7.00\arcmin & I & 27.2 & 17 & 12.3& $25.0\times25.0$\\
3C 31 & 44.19\arcmin & I & 51.8 & 95 & 10.5& $45.6\times45.6$\\
3C 310 & 5.09\arcmin & I & 182.7 & 15 & 10.2& $26.1\times26.1$\\
3C 326 & 20.15\arcmin & II & 47.6 & 14 & 41.1& $26.2\times26.2$\\
3C 33 & 4.15\arcmin & II & 118.1 & 68 & 13.4&  $25.0\times25.0$\\
3C 382 & 3.20\arcmin & II & 63.4 & 53 & 9.8& $30.6\times30.6$\\
3C 442a & 10.08\arcmin & II$^{\dagger}$ & 53.3 & 31 & 13.4& $25.0\times25.0$\\
3C 449 & 24.97\arcmin & I & 28.7 & 9 & 14.0& $46.6\times46.6$\\
3C 452 & 4.53\arcmin & II & 165.0 & 50 & 8.6& $46.6\times46.6$\\
3C 457 & 3.49\arcmin & II & 41.3 & 35 & 12.2& $25.0\times25.0$\\
3C 46 & 2.63\arcmin & II & 28.6 & 57 & 6.5& $38.5\times38.5$\\
3C 465 & 9.75\arcmin & I &  110.0 & 17 & 12.0& $25.0\times25.0$\\
3C 76.1 & 3.32\arcmin & I & 23.8 & 17 & 11.2&$25.0\times25.0$\\
3C 98 & 5.11\arcmin & II & 121.7 & 60 & 10.1&$25.0\times25.0$\\
4C 12.03 & 3.59\arcmin & II & 27.3 & 5 & 12.7&$25.0\times25.0$\\
NGC 6109 & 14.74\arcmin & I & 27.6 & 16 & 18.8& $31.3\times31.3$\\
NGC 7385 & 13.83\arcmin & I & 37.7 & 30 & 12.8 & $25.0\times25.0$\\
 &&&&&&\\
{\bf Poor LBA quality } &&&&&&\\
\hline
	3C 293 & 4.28\arcmin & I &  31.4 & 125 & 8.8&\\
	3C 314.1 & 3.35\arcmin & I &  37.35& 33 & 39.91&\\
	3C 315 & 3.36\arcmin & I &  62.2 & 15 & 51.6&\\
	3C 321 & 5.13\arcmin & II &  45.4 & 118 & 8.8&\\
\hline
	\end{tabular}
	\tablefoot{Of the sources at the bottom of the table, 3C~314.1 and 3C~315 was excluded from the analysis due to poor quality of the LBA image (calibration errors). 3C~293 and 3C~321 were excluded because there is no visible two sided lobes in this source, only the core and hotspots at the end of the lobes.\newline
	$^{\dagger}$ In the reference, this source is listed as FR~I, but the colour-colour diagram is more consistent with an FR~II. See Section~\ref{fig:injection_index} for a discussion.}
	\tablebib{(a) ~\url{https://3crr.extragalactic.info/}}
\end{table*}

\begin{figure}[th]
	\centering
	\includegraphics[width=1.\linewidth, trim={0.cm 0.cm 0.cm 0.cm},clip]{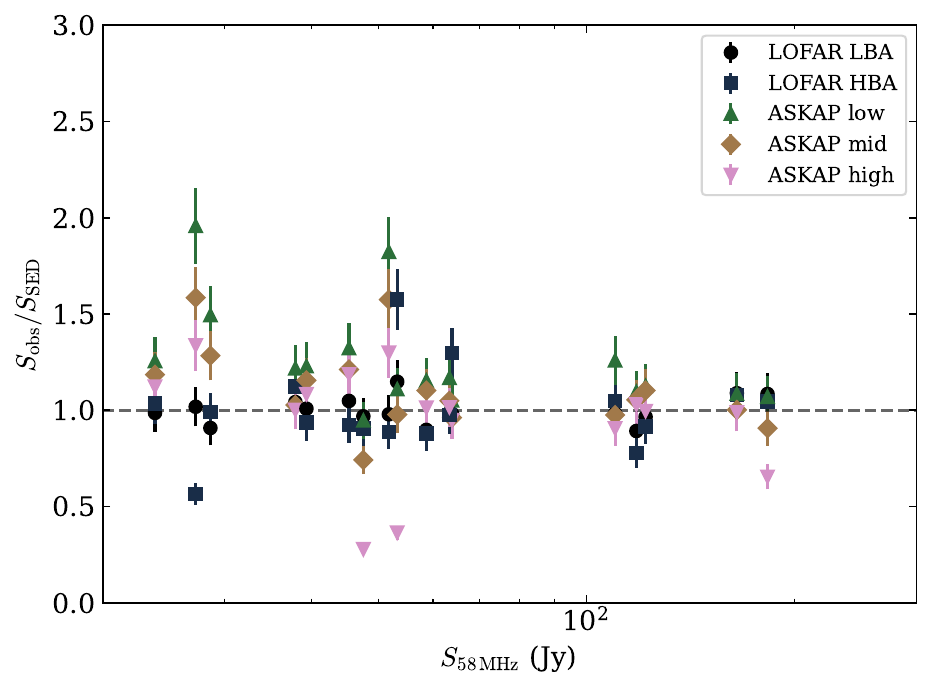}
	\caption{Ratio of measured integrated flux density over the expected flux density from the SEDs from \citet{Boxelaarprep} for all frequency maps as a function of the LBA flux density.}
	\label{fig:flux_scale_check}
\end{figure}

Table~\ref{tab:the_subsample} shows the subsample and summarises some characteristics. The final beam refers to the resolution of the images used for the analysis in Section~\ref{sec:spidx}. As the beam of RACS maps tends to be quite elongated, due to the low elevation of most sources for RACS, we were forced to convolve with a relatively large beam ranging from $25\arcsec\times25\arcsec$ to $46\arcsec\times46\arcsec$. This is mainly due to the low resolution of RACS-low, but as it is an essential frequency to cover between LoTSS and GHz regime, we did not discard it. For the LoTSS DR3 \citep{2026A&A...707A.198S} 144~MHz cutouts we used the 20\arcsec\ maps (instead of the higher resolution 6\arcsec\ variation) to be more sensitive to diffuse emission and since the poorer resolution of RACS-low does not compromise image quality. The resulting subsample containing 22 sources is summarised in Table~\ref{tab:the_subsample}. To the authors' knowledge, there are only two sources in this subsample which have been studied in the ultra-low frequency regime before. 3C~31  and 3C~223 are the only ones previously studied using LBA \citep{Heesen2018, 2016MNRAS.458.4443H} and 3C~223 has also been studied at 74~MHz using the VLA \citep{Orru2010}. This emphasises the novelty of our ultra-low frequency maps. Notable extended sources at $\text{DEC}>40\degr$ missing from this list due to the lack of RACS data are e.g. 3C~35, 3C~83.1B, DA~240, 4C~73.08 and NGC~6251. The Perseus cluster (3C~83.1B \& 3C~84) has also been observed with broadband observations by \citet{2026A&A...709A..64G}. 

\begin{figure*}[th]
	\centering
	\includegraphics[width=.335\linewidth, trim={0.cm 1.2cm .cm 0.cm},clip]{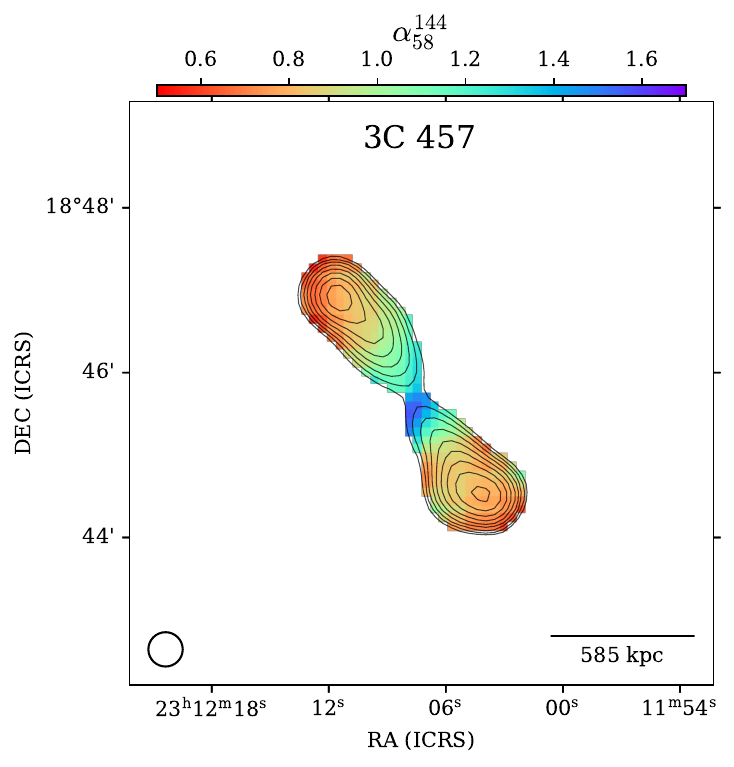} 
	\includegraphics[width=.32\linewidth, trim={.7cm 1.2cm 0.cm 0.cm},clip]{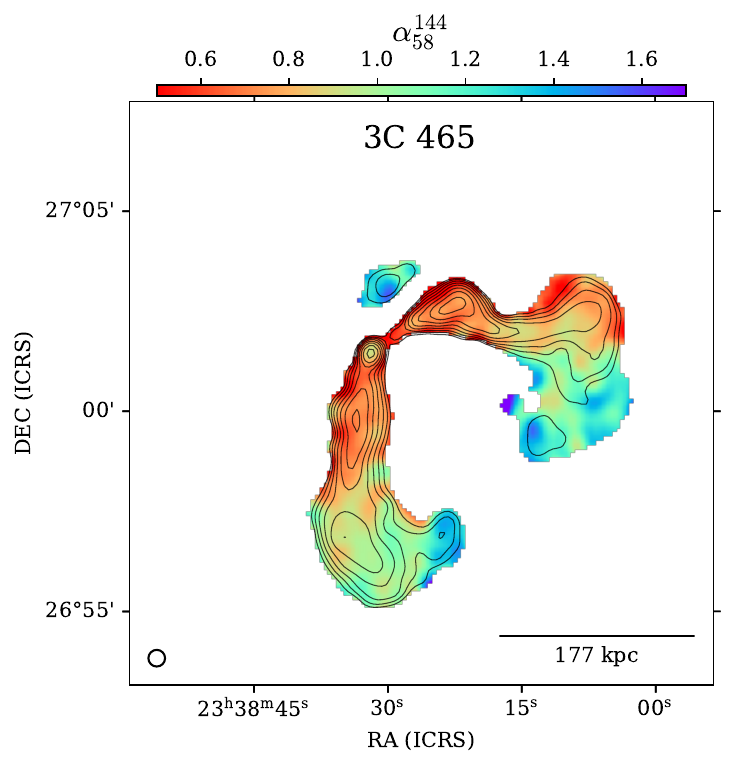} 
	\includegraphics[width=.32\linewidth, trim={.7cm 1.2cm 0.cm .cm},clip]{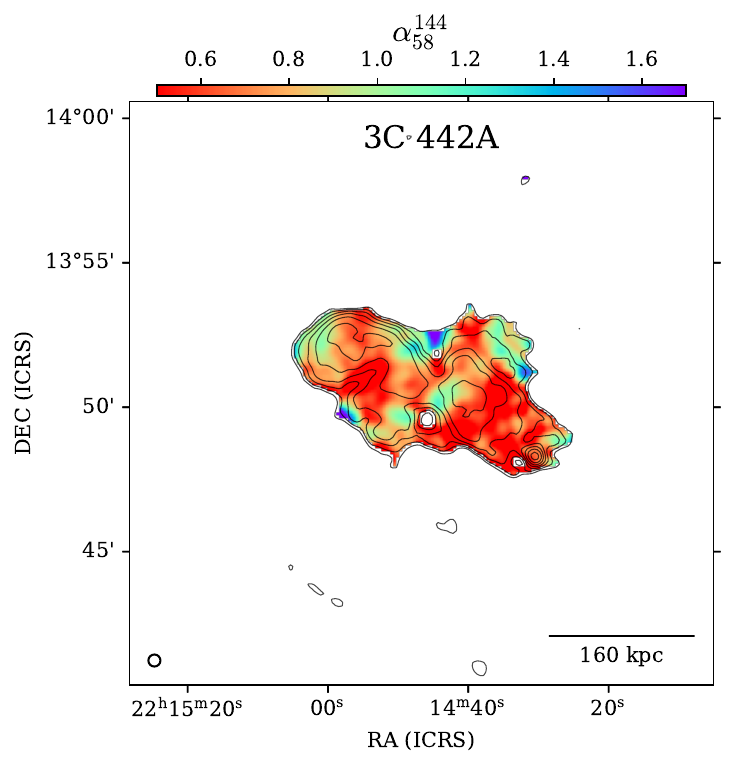} \linebreak
	\includegraphics[width=.335\linewidth, trim={.cm 0.cm 0.cm 0.cm},clip]{Assets/spidx/spidx_3c457__0,1_.pdf} 
	\includegraphics[width=.32\linewidth, trim={.7cm .cm 0.cm .cm},clip]{Assets/spidx/spidx_3c465__0,1_.pdf} 
	\includegraphics[width=.32\linewidth, trim={.7cm .cm 0.cm .cm},clip]{Assets/spidx/spidx_3c442a__0,1_.pdf} 
	\caption{Spectral index maps for 3C~457, 3C~465, and 3C~442A at $25\arcsec\times25\arcsec$. The colour scale is set between 0.5 and 1.7 for every image. The contours are drawn at $[3, 5, 10, 50, 100, 200]\times \sigma_{\text{rms}}$ of the LBA image. Every map shows spectral index values with a threshold of $2\sigma_{\text{rms}}$. Finally, the beam for that map and the physical scale of the source are also indicated.}
	\label{fig:spidx_ex}
\end{figure*}

\subsection{Spectral index maps}
In order to study the radio spectral shape in Section~\ref{sec:colour-colour} we need good spectral index maps. We constructed spectral index maps for all the extended sources by first convolving all images to a minimum common circular beam and regridding those images to equal world coordinate system projections. The astrometric alignment is done in an automated way based on the locations of various point sources in the field around the source of interest. Depending on the field and map noise there might not be many point sources bright enough to align with. Therefore in some cases additional manual, sub-pixel alignment was needed to obtain usable spectral index maps. We checked for flux density offsets by plotting integrated flux densities compared to expectations from the SEDs of \citet{Boxelaarprep} in Figure~\ref{fig:flux_scale_check}. Although we did not find significant offsets, this check did show that maps exhibit a larger scatter for low surface brightness sources, which will result in poor quality spectral index maps and colour-colour plots.

Uncertainties on spectral indices derived from integrated regions or single pixels may be divided into two different contributions. Uncertainties on the integrated flux densities are a combination of stochastic errors from underlying rms noise, independently distributed per beam area, and the systematic calibration error, which effects the entire image, or at least the source of interest, equally. For the purpose of interpreting the colour-colour diagrams in the next subsection it is useful to keep these uncertainty contributions separated, since one of the contributions is a systematic error that gives rise to a systematic shift rather than scatter in the colour-colour diagram. The total combined uncertainty on measured flux density is
\begin{align}
    \label{eq:delta_s}
	&&\Delta S_{\nu} =\sqrt{\left(J\sigma_{\text{rms}} \sqrt{\frac{n_{\text{reg}}}{A_{\text{beam}}}}\right)^2 + \Delta S_{\text{cal}}^2},
\end{align}
where $J=3$ is the on-source noise multiplier as also used in spectral age fitting software \texttt{BRATS} \citep{Harwood2013MNRAS.435.3353H}, $n_{\text{reg}}$ is the number of pixels in the region over which the flux density is measured and $A_{\text{beam}}$ is the beam area in pixels. $\Delta S_{\text{cal}}$ is the calibration error which normally takes the form $E_{i}S_{\nu}$ where $E_i$ is a percentage and $S_{\nu}$ the measured flux density. Recently, \citet{Boxelaarprep} determined the calibration error for LBA to be 10\%. For LoTSS DR2, the conservative estimate is 10\% \citep{Shimwell2019}, which we used here. For RACS: $\Delta S_{\text{cal}}= 0.05\text{ mJy} + 0.07S_{\nu}$ \citep{RACS2020}. As we only considered two frequencies at a time the spectral index is simply given by $\alpha = (\log S_1 -\log S_2)/(\log \nu_1 -\log \nu_2)$ with uncertainty 
\begin{align}
	&& \sigma_{\alpha}=\log^{-1}\left(\frac{\nu_1}{\nu_2}\right)\sqrt{\left(\frac{\Delta S_1}{S_1}\right)^2 + \left(\frac{\Delta S_2}{S_2}\right)^2},
	\label{eq:spidx_err}
\end{align}
where $\Delta S_i$ refers to both the calibration or the flux density measurement error of Equation~\ref{eq:delta_s}. 

Figure~\ref{fig:spidx_ex} shows examples of the low- and high-frequency spectral index maps for three archetypic sources. FR~II galaxy 3C~457, FR~I galaxy 3C~465 and the FR~II galaxy 3C~442A with a more irregular morphology. 3C~457 shows hotspots with low spectral index ($\sim0.6$) at very low frequencies which steepen towards the AGN. Due to the inverted spectrum of the central AGN, the LOFAR spectral index map does not detect it, while the contribution from the core is visible in the high frequency map. As this emission is distinct from the lobe and hotspot emission, the central regions are ignored when constructing the colour-colour diagrams. The spectral index maps of 3C~465 highlight the value of the 50~MHz LOFAR LBA observations. The LBA is sensitive to the emission far away from the core and we were able to measure the spectral index in those regions while this is not possible using the higher frequency data. 3C~442A shows something similar: the LOFAR spectral index map reveals additional diffuse emission perpendicular to the bright east-west oriented lobe. The spectral index is relatively constant throughout the source compared to the other two sources in the figure.

\section{Results}
\label{sec:spidx}
Throughout the remainder of this work, the injection index $\injindx$ refers to the spectral index of the freshly injected cosmic ray population, defined through the initial power-law energy distribution $N(E) \propto E^{-\delta}$ with $\delta = 2\injindx + 1$. It represents the flattest spectral index expected anywhere in the source, corresponding to zero-age plasma, and is one of the key parameters of the JP and KP ageing models described in Section~\ref{sec:introduction}. Its measurement from colour-colour diagrams is described in detail in Section~\ref{sec:injidx}.

Both low frequency spectral index maps ($\alpha^{144}_{58}$) and high frequency spectral index maps ($\alpha^{1367}_{887}$) are shown in Appendix~\ref{app:spidx_maps_all}. The uncertainty on the spectral indices can be derived from the error bars in Appendix~\ref{app:colour-colour-diagram}. 

\subsection{FR~Is}
The overall low frequency spectral index $\alpha^{144}_{58}$ in the lobes of FR~I type radio galaxies varies from 0.5 close to the core to above 1.0 far into the lobes. There is a great variety of spectral indices depending on the location in the lobe. Many FR~Is have a spectral index of about 0.5 near the core, consistent with first-order Fermi acceleration. In Section~\ref{sec:injidx} we showed that this value is consistent with most measured $\injindx$, which implies that the emission close to the core is representing the injected plasma. The spectral index maps of this class of radio galaxies are generally more complex than for the FR~IIs due to the low surface brightness in the far ends of the diffuse lobes. 

\subsection{FR~IIs}
The global low frequency spectral indices in the lobes and, for FR~IIs, the hotspots are listed in Table~\ref{tab:results}. The low frequency spectral index in the hotspots vary between 0.5 and about 0.8, with values reaching up to 0.9 for 3C~192 and 3C~382. Values between 0.7 and 0.8 are not uncommon for higher frequency observations \citep{Harwood2013MNRAS.435.3353H, Harwood2015MNRAS.454.3403H}. Furthermore, \citet{2016MNRAS.458.4443H} has shown that hotspot spectra can remain steep to ultra-low frequencies. Table~\ref{tab:results} reveals the spectral index in the hotspot is steeper than the measured injection index in almost every case. This was hinted at by \citet{2017MNRAS.466.2888H} for 3C~223 and 3C~452 using 144~MHz as their lowest frequency. Here we confirmed this observation for a larger sample while also including 58~MHz observations.  The colour-colour diagrams of 3C~452, 3C~457, 3C~46 and 3C~284 (see Figure~\ref{fig:colour-colour_ex} for 3C~457, and Figure~\ref{fig:colour-colour-all-frii} for the other sources) indeed show the hotspot regions (in red) never reach the solid line where we expect freshly injected electrons. This can be due to absorption or more complex acceleration mechanisms within the hotspots or mixing. In Section~\ref{sec:discussion} we discuss the topic of mixing in more detail. \citet{2017MNRAS.466.2888H} found a $\injindx$ of 0.68 for 3C~223 and 0.70 for 3C~452 through a grid search of the JP parameter space. While we found a similar value for 3C~452, we obtained a very flat $\injindx=0.46\pm0.06$ for 3C~223. The discrepancy can be due to the strong steepening at higher frequencies, driving up the $\injindx$ measurement if ultra-low frequency data is missing.  

\subsection{Colour-colour diagram}
\label{sec:colour-colour}
The synchrotron colour-colour diagram is a way of examining the spectral shape of AGN lobes and other radio sources. It was first proposed by \citet{Katz-Stone1993} to characterise the spectral properties of Cygnus~A. It has since then been used in multiple works as a way to fit ageing models to FR~I and II type sources and radio relics \citep[e.g.][]{2012A&A...546A.124V, Rajpurohit2020, bruno2024, Lusetti2024, Rajpurohit2025, 2026A&A...707A..49D}. The colour-colour diagram is defined by the plane of two spectral indices between either three or four observing frequencies. In this work those could be either $\alpha_{144}^{57}$ or $\alpha^{57}_{887}$ against $\alpha^{887}_{1367}$. The diagram provides visual information on the spectral shape of the lobe emission and the time evolution thereof. The line $\alpha^1_2=\alpha^3_4$ defines the pure power-law with everything below it showing a steepening at higher frequency, and every point above the line showing a concave spectrum. Any line parallel to it encodes a spectrum of constant curvature. 

Ageing models trace out curves in this diagram, where each point along a curve corresponds to a particular plasma age \citep[see][for details]{Katz-Stone1993}. The JP model produces a particularly simple curve: it follows the power-law line up to the injection index $\injindx$, then diverges steeply downwards. Only the injection index affects the shape of the curve (it sets the point along the power-law at which the curve begins) while the magnetic field strength only controls how quickly a source moves along the curve with increasing age.

A colour-colour diagram is particularly useful to obtain information on the global characteristics of the source. The single diagram contains both information about the injection index and how closely the source follows pure ageing models. Since all regions of the lobe are in one parameter plane this representation is more compact than e.g. point-by-point age fitting. In analysis involving age fitting, the magnetic field strength and age of the plasma are almost impossible to model accurately using just radio maps since the two parameters are fully degenerate. The degeneracy can be ignored in the colour-colour diagram as the curve of an ageing model is independent of magnetic field and age. The magnetic field only changes how 'quickly' a point moves along the curve per age interval. The slope on the diagram is inherent to the model and independent of any parameter. Only the injection index can tune its shape by translating its starting point along the power-law line.

\begin{figure*}[th]
	\centering
	\includegraphics[width=.34\linewidth, trim={0.cm 0.cm 0.cm 0.cm},clip]{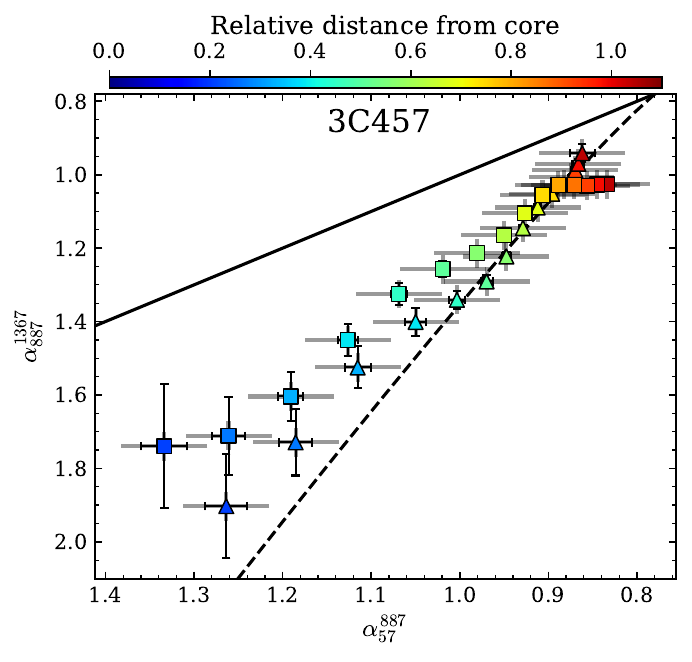}
	\includegraphics[width=.315\linewidth, trim={1.cm 0.cm 0.cm 0.cm},clip]{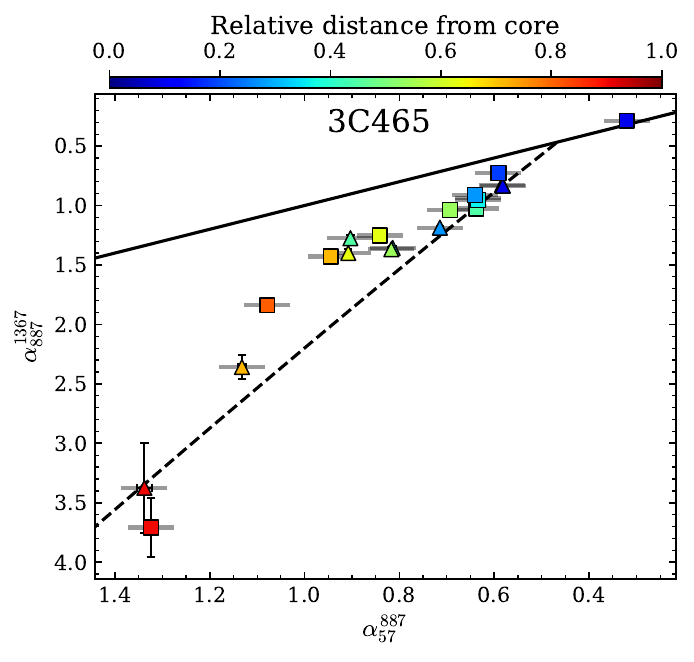}
    \includegraphics[width=.32\linewidth, trim={1.cm 0.cm 0.cm 0.cm},clip]{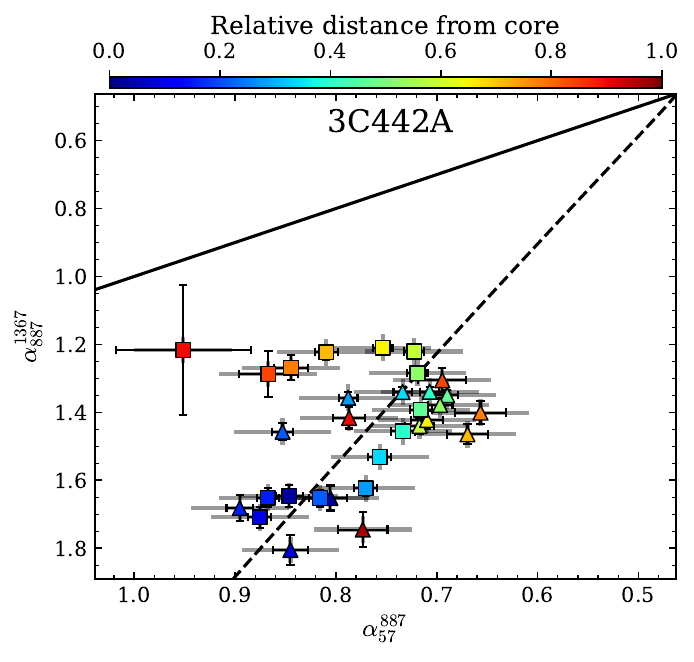}
	\caption{Colour-colour diagrams for a typical FR~II galaxy 3C~457 (left), a typical FR~I galaxy 3C~465 (middle), and 3C~442A (right) in the $\alpha^{57}_{144}-\alpha^{887}_{1637}$ plane. The colour scale denotes the distance from the core relative to the approximate size in Table~\ref{tab:the_subsample}. The solid grey error bar denotes the possible systematic shift of data points due to calibration errors and the thin black error bars denote the rms flux uncertainty. The solid line denotes a pure power-law and the dashed line a JP model with $\alpha_{inj}=0.78\pm0.09$ for 3C~457, $\alpha_{inj}=0.5\pm0.1$ for 3C~465 and $\alpha_{inj}=0.46\pm0.09$ for 3C~442A.} The two lobes are differentiated by either showing a square or triangle. For both sources the region containing the core is excluded from the figure.
	\label{fig:colour-colour_ex}
\end{figure*}

Figure~\ref{fig:colour-colour_ex} shows the colour-colour diagrams of the same sources from Figure~\ref{fig:spidx_ex}. To retain homogeneity across different sources, the colour coding in Figure~\ref{fig:colour-colour_ex} is the relative distance from the core where 1 denotes half the size from Table~\ref{tab:the_subsample}. In other words, the approximate maximum extent of the source relative to the core. Because of this definition, source 3C~236 has points exceeding one because it has asymmetric jets. The bold grey error bars denote the calibration error and are, due to the $\Delta S/S$ dependence in Eq.~\ref{eq:spidx_err}, equal for all data points in the diagram. The thin black error bars denote the rms uncertainty on the flux density measurement which depends on the signal to noise at the data point. The distinction makes clear which points in the diagram are from regions with very low surface brightness emission. Note that the grey error bars do not denote a $1\sigma$ uncertainty but rather the possible extent of any systematic offset due to the calibration error. This offset should affect all data points equally. The colour-colour diagrams for all sources in the sample are shown in figures~\ref{fig:colour-colour-all-fri} and \ref{fig:colour-colour-all-frii}. 

We adopted two strategies for obtaining the spectral index across the lobes. For morphologically straight sources (e.g. 3C~457), we regridded the image while rotating it to be aligned on the image x-axis and integrated the flux in regular intervals across the source. For sources that do not behave this way (see e.g. Figure~\ref{fig:3c465-multi-component} for regions of 3C~465) we placed circular regions along the lobe ridge lines. Four of the galaxies (see e.g. 3C~76.1, 3C~310, 3C~382 and 3C~442A in Appendix~\ref{app:spidx_maps_all}) have 'fat'\footnote{With a fat lobed source we mean e.g. 3C~76.1, which is not morphologically elongated.} lobes with no clear ridge line and the colour coding for those sources is therefore somewhat arbitrary.

From the colour-colour diagrams we pointed out several aspects. First, there is a large portion of sources where the colour-colour diagram is not well behaved, in the sense that they do not conform to expectations from previous work including \citet{Katz-Stone1993}, i.e. a slim locus of points parameterised by the distance from the core. These are generally the sources with atypical lobe structure like fat-double radio galaxies. An example is 3C~310 which is known for its complex lobe structure in the form of a radio ring  in the southern lobe \citep{Gizani2002}, which is unresolved at 15\arcsec. Factors like the external environment and jet kinematics seem to influence the jets in such a way that it can no longer be described by a simple ageing model in a colour-colour diagram at the current resolution.

For this particular case, 6\arcsec\ LoTSS observations resolve the finer structures in 3C~310. It is clear the 15\arcsec\ LBA data is not sufficient. The low resolution and poor fidelity emphasise the need for higher resolution broad-band imaging with LBA, see the discussion for an outlook on this subject. This behaviour also results in high uncertainties on $\injindx$ measurements which we discuss in Section~\ref{sec:injidx}.

Secondly, we note the locus of points does not reach the zero age power-law for FR~II galaxies. This is telling us that we did not observe zero age emission in the hotspots, but instead either an already aged spectrum or a spectrum with a low frequency energy cut-off as suggested by \citet{Katz-Stone1993}. 3C~284, 3C~457 and 3C~46 show signs of this behaviour. The colour-colour diagram of 3C~223 shows good agreement with JP but the hotspots are very far from zero age. This could again be due to the low resolution of our images where at the hotspot regions we sumed over both hotspot and surrounding emission which will seemingly result in an aged spectrum. 

In well-behaved sources, the colour-colour diagrams reveal a clear and systematic difference between FR I and FR II galaxies. For FR IIs, the data points closest to the hotspots lie nearest to the zero-age power-law line, with spectral steepening increasing monotonically towards the core, tracing the backflow of aged plasma away from the injection site. For FR Is, the behaviour is reversed: the flattest spectra are found closest to the core, with progressive steepening towards the outer lobes.

Finally, we saw that the locus of points has less scatter in the $\alpha^{887}_{57} - \alpha^{1367}_{887}$ plane compared to the $\alpha^{144}_{57} - \alpha^{1367}_{887}$ plane. This is due to the larger frequency difference in the former. In the colour-colour diagrams we therefore plotted  $\alpha ^{1367}_{887}$ against $\alpha ^{887}_{57}$ in favour of $\alpha^{144}_{57}$ due to the fact that the uncertainty due to flux density scale offsets is larger for smaller frequency intervals. The shift $\Delta\alpha$ in spectral index due to the relative flux density offsets $\delta$ of two flux density measurements is 

\begin{equation}
    \Delta\alpha = \frac{\log{\delta_1/\delta_2}}{\log{\nu_1/\nu_2}}.
\end{equation}
For LOFAR LBA and HBA, for which $\delta$ is in between 0.9 and 1.1 given the flux density scale uncertainties (see Table~\ref{tab-all-obs-new}), the shift in spectral index can be up to 0.2, while for LBA and ASKAP-low this offset can be only up to 0.06.

\subsection{Measuring injection indices}
\label{sec:injidx}
We measured the injection index from the data points in the colour-colour diagram. The expected flux density at a certain frequency for an electron population of a certain spectroscopic age $t$ in a magnetic field $B$ and injection index $\injindx$ is given by Equation~\ref{eq:int}. Since the age and magnetic field are degenerate parameters in the JP model, we fixed the magnetic field and let the age and injection index be the only relevant varying parameters. The magnetic field strength is set to $B=B_{\text{CMB}}/\sqrt{3}$ for the entire source as it was shown by \cite{Hardcastle2013MNRAS.433.3364H} that replacing $B$ by a more physical Maxwell-Boltzmann distribution of the magnetic field does not significantly change the mean emissivity of the model.

Given a measured spectral index $\alpha_{12}$ at frequencies $\nu_i$ and $\nu_j$, flux densities (from which we derived the injection index) are modelled by $S_{\nu_i}(t,\alpha_\text{inj})$ and $S_{\nu_j}(t,\alpha_\text{inj})$. We can summarise this by saying that the spectral index for a given frequency pair is given by $\alpha^i_j(t, \injindx)$. The location of a measurement in the colour-colour diagram depends on two of these spectral indices, e.g. $\alpha^i_j$ and $\alpha_k^l$. To find the injection index corresponding to every measurement we numerically found the values of $t$ and $\injindx$ for which $\alpha^i_j$ and $\alpha_k^l$ correspond to the measured spectral indices. 

We know that there is only one unique solution for the $t$, $\injindx$ pair because we qualitatively know $\alpha^i_j$ is a strictly increasing function of both $t$ and $\injindx$ (both an older plasma and a higher injection index steepens the spectral index). This means, given a value of $\alpha^i_j$, there is at most one $t$ for each $\injindx$, hence a strictly increasing or decreasing function in the $(t,\injindx)$ plane. So, there can be at most one set of parameters for which the JP model goes through a point in the colour-colour diagram. 

We found these sets of parameters using minimisation algorithms since the integral in Eq.~\ref{eq:int} cannot be inverted analytically. The best fitting injection index ($\bar{\alpha}_{\text{inj}}$) for the entire source is then found by root mean square minimisation. The uncertainty on the obtained injection index is given by the root mean squared deviation (RMSD)
\begin{equation}
	\alpha_{\text{inj, RMSD}}=\sqrt{\frac1N\sum_i^N (\alpha_{\text{inj},i}-\bar{\alpha}_{\text{inj}})^2}.
\end{equation}

\begin{figure*}[th]
	\centering
	\includegraphics[width=.955\columnwidth, trim={.cm .cm .cm .cm},clip]{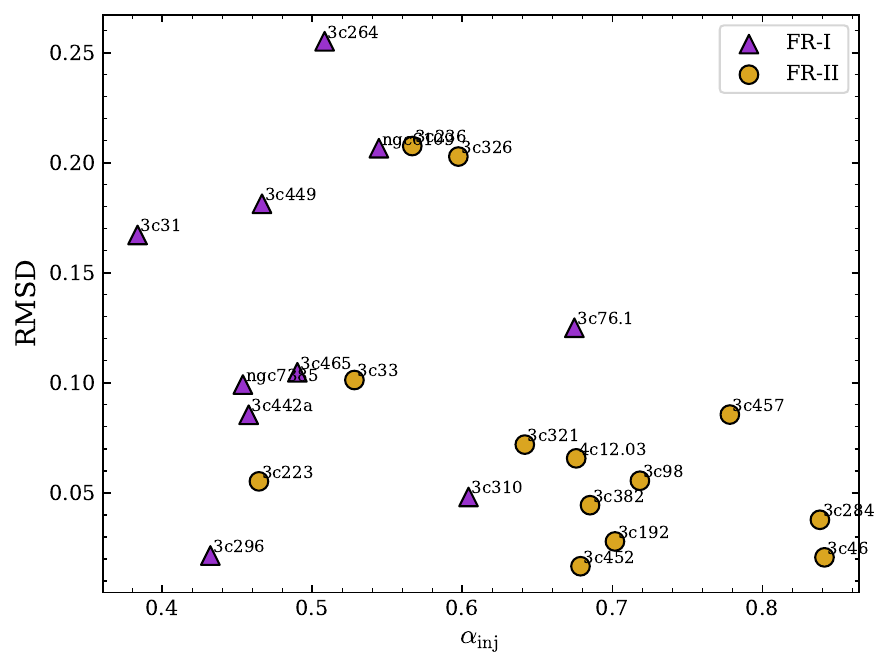} 
	\includegraphics[width=.93\columnwidth, trim={.cm .cm .cm .cm},clip]{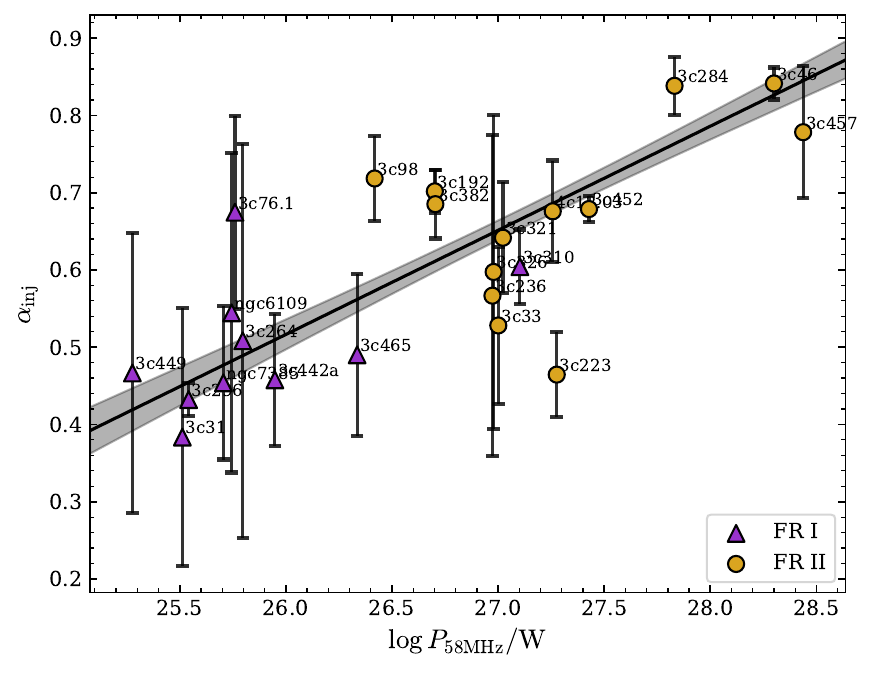} 
	\caption{Left panel: Measured injection index plotted against their uncertainty from RMSD where we differentiated between FR~I and FR~II type galaxies. Right panel: shows the apparent relation between the  measured injection index and the total radio power. Purple triangles depict the FR~Is in our sample and the orange circles depict the FR~IIs. The black line and shaded area are a linear MCMC fit and the uncertainty respectively.}
	\label{fig:injection_index}
\end{figure*}

The measured injection indices and the RMSD uncertainties are shown in the left panel of Figure~\ref{fig:injection_index}. The right panel of this figure shows the measured injection index against the total radio power of the source and hints at a relation between the total radio power and $\injindx$. Since the correlation, characterised by a correlation coefficient\footnote{When a correlation coefficient is mentioned, we refer to the Pearson correlation coefficient.} of 0.75, is significant, we fitted a linear function in the $\injindx-\log{P}$ plane using a bivariate correlated errors and intrinsic scatter \citep[BCES;][]{1996ApJ...470..706A} method implemented by \citet{2012Sci...338.1445N} and find $\injindx=0.135^{+0.014}_{-0.013}\log{P} -3.00^{+0.36}_{-0.35}$. The separation of the two FR types is driven mostly by luminosity \citep[noted already by][]{Fanaroff1974} but is enhanced by the injection index. In general, we observed FR~I type galaxies to have a less well behaved colour-colour diagram compared to FR~IIs. 
In Figure~\ref{fig:colour-colour_ex} however, the colour-colour diagram of 3C~442A shows more erratic behaviour near the source outskirts compared to other FR~II sources. This could be attributed to missing flux in the ASKAP maps for the low surface brightness emission. Contrary to the literature cited in Table~\ref{tab:the_subsample}, the colour-colour diagram shows to be more consistent with a FR~II type galaxy. The distance to the core is inversely proportional to the deviation from the perfect power-law spectrum.

We suggest that environmental effects can have significant influence on the lobe plasma and induce local variations due to shocks and turbulence; this can be the case for e.g. 3C~264, 3C~465 and NGC~6109 which reside in dense cluster environments. Also re-acceleration can play a role in disturbing expected behaviour in colour-colour diagrams (See Section~\ref{sec:reacceleration}).

We did not observe significant differences between the two lobes when measuring the injection index, meaning that material in opposite hotspots and jets are similarly accelerated. Though for 3C~284 we did observe significantly different spectral index for the two lobes, 0.80 for the west lobe and 0.84 for the east lobe. In the radio maps the lobes are asymmetric, with the east lobe being brighter than the west lobe due to projection and beaming effects.

The luminosity-injection index relation has previously been reported by \citet{2008MNRAS.385.1286J} in a sample of FR~II galaxies. Our results seem to suggest that the relation continues down to the FR~I regime. However, the correlation coefficient for FR~Is only is just 0.4 and considering the uncertainties, the injection index for this type of object does not correlate with the total radio power and the injection index is $\sim0.50\pm0.08$. This result is fairly consistent with, though slightly lower than, spectral indices often quoted in the literature, such as 0.55 by \citet{2005ApJ...626..748Y}.  For FR~IIs, the correlation is stronger and they tend to have higher injection indices for higher total radio luminosities. The injection indices are typically $\injindx\gtrsim0.6$ with values up to 0.85 for this sample. Only 3C~223 has a spectral index slightly lower at $\injindx=0.46\pm0.06$. Since at injection losses are  negligible, the injection index should be characteristic of the acceleration mechanism. For example, first-order Fermi acceleration in strong, non-relativistic shocks \citep{1978MNRAS.182..147B,1978ApJ...221L..29B} has $\injindx=0.5$. Our measurements for FR~I galaxies seem consistent with this mechanism, although, due to the high uncertainty on the injection index, we cannot make strong claims about possible (re)acceleration mechanisms with the current data.

\section{Discussion}
\label{sec:discussion}
The analysis in this work is largely based on spectral index measurements. The main limitations of any spectral index measurement are flux scale errors of the used images, which are 10\% for the observations at the lowest two frequencies. This, combined with the fact that the HBA frequency is only 2.5 times higher than the LBA frequency, can lead to systematic offsets of the spectral index of up to 0.2, which is significant for spectral index measurements down to 0.4. Another limitation for this work was the sometimes poor quality of the LBA data. This is due to the fact that the data have a limited bandwidth of only 3~MHz which leads to high RMS noise values and low signal to noise especially for regions with low surface brightness. The narrow bandwidth furthermore results in relatively poor uv-plane coverage and image fidelity. Despite these challenges we were able to produce spectral index maps below 150~MHz for 22 3CRR sources and estimate injection indices for them.  

The colour-colour diagrams show a clear distinction between FR~Is, showing increasing spectral steepening away from the core, and FR~II galaxies, showing a backflow of aged plasma from the hotspots towards the core. This is consistent with the established physical picture of the two classes \citep[e.g.][]{1999A&A...344....7P, 2014MNRAS.437.3405L, Harwood2013MNRAS.435.3353H, Brienza2020, 2026arXiv260412043R}. In FR~IIs, particle acceleration occurs predominantly at the hotspots, where the jet terminates in a strong shock. Electrons then flow back into the lobes, losing energy through synchrotron and inverse Compton radiation as they travel away from the hotspot, producing the observed gradient of spectral steepening from the hotspot inwards. In FR Is, in contrast, there are no hotspots; acceleration is believed to occur in the inner jet, close to the AGN, in the region where the jet decelerates and dissipates. Electrons then propagate outwards into the lobes, ageing progressively as they do so, which naturally produces spectral steepening with increasing distance from the core. Our measurements show that the low-frequency spectral steepening is less pronounced in FR~Is, suggesting that synchrotron losses are not the driving factor in the spectral shape of this class of sources. The inclusion of 58~MHz data is particularly valuable here, as the ultra-low frequency spectral index is sensitive to the oldest electron populations and therefore provides a cleaner separation between injection and ageing signatures than is possible with higher frequency data alone. The colour-colour diagram at these frequencies thus provides a spatially resolved, largely model-independent confirmation of the FR~I and FR~II acceleration dichotomy.

\subsection{Effects of mixing and complex spectral structures} 
One has to be careful of over-interpreting these diagrams since they are sensitive to shifts in flux scale and accuracy of the image alignment. The scaling of one of the considered flux densities for a given shift in a spectral index on the colour-colour plane becomes larger for larger ranges in frequency. For example, a shift in $\alpha^{144}_{57}$ of 0.1 corresponds to a flux scaling of $S_{57\text{ MHz}}$ of 1.1, while the shift for a much larger frequency jump $\alpha^{887}_{57}$ of 0.1 the scaling of $S_{57\text{ MHz}}$ is 1.3. Thus, the sensitivity to flux density uncertainties decreases for larger frequency intervals. Moreover, since we are dealing with a sub-optimal resolution of 25\arcsec\ there is likely mixing of electron populations within an observing beam especially for the least extended sources. There may also be multiple physical effects manifesting in the lobes that leave signatures in the diagram that are hard to separate. This can be changing magnetic field strengths, re-acceleration or changes in injection index along the lobe and to the hotspots related to the accretion history of the AGN. 

\begin{figure}[th]
	\centering
	\includegraphics[width=1.\columnwidth, trim={.4cm 0.2cm 1.5cm 1.2cm},clip]{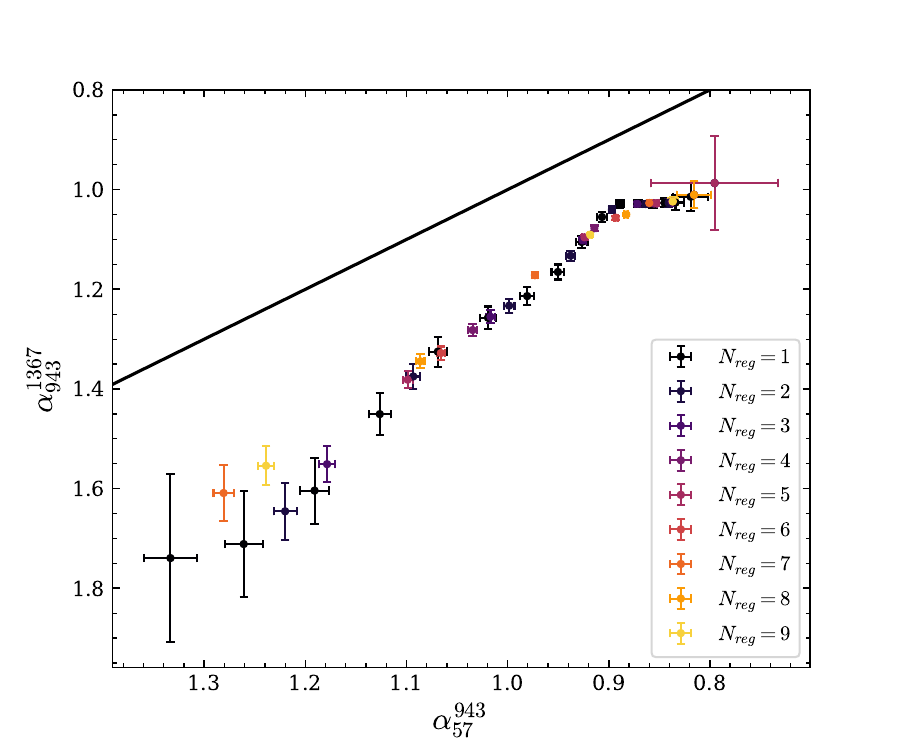} 
	\caption{Colour-colour diagram of 3C~457 to visualise the effect of mixing. The different colours indicate the measured spectral indices for different region sizes in pixels.}
	\label{fig:mixing}
\end{figure}

Since it is possible that the magnetic field changes along the lobes it is challenging to measure the source age. Due to the low resolution there will be mixing of electron populations of different ages. We tested the influence of mixing on the measurement of the $\injindx$ by measuring the injection index from the colour-colour plot using differently sized regions of the image. Using larger regions allows more electron population mixing. Figure~\ref{fig:mixing} shows the colour-colour diagram for different region sizes. 

The figure shows that depending on the regions, the data points translate on the diagram due to an effectively different measurement of the magnetic field. It is important to note that the points all stay on the same trajectory, meaning that the measurement of  $\injindx$ is not affected by the increased mixing of electrons of different ages or different magnetic field environments. If the injection index were influenced we would expect to observe a shift of the trajectory with respect to the solid power-law line. From this we can conclude that we are measuring the same value for the injection index as we would with higher resolution data.

It has been shown that some FR~I galaxies show a two-component spectrum with a flat and steep component \citep[see e.g.][]{1997ApJ...488..146K,1999A&A...349..381H}. Often this is characterised by a collimated jet surrounded by a more diffuse 'sheath' with a steeper spectrum. Even in resolved maps, some lobe regions exhibit this two-component spectrum, effectively mixing the components when trying to measure spectral index slopes. Since we assumed the lobes to be consistent with just a single component in estimating the injection index, the existence of multi-component lobes may, at first sight, slightly complicate the spectral modelling we carried out in this work. 

This is not likely to be the case however, since, for colour-colour diagrams that are not well behaved, which is the majority of them, we cannot make any claims about the existence of multiple components and the estimate of the injection index will in any case have a high uncertainty. If, on the other hand, a colour-colour diagram is already fairly well behaved, and the lobes do not have multiple components everywhere, it is still possible to get accurate estimates for the low frequency injection index as we averaged over the entire source to find this index. We showed that the trend in a colour-colour diagram  does not necessarily depend on the resolution, it is more likely related to the telescope sensitivity and source morphology.

Furthermore, if the diagram is well behaved it could be possible to identify the existence of a multi-component lobe.  One of the more well behaved colour-colour diagrams, 3C~465, shows hints of a double-component lobe (See Figure~\ref{fig:colour-colour_ex}). The regions close to the core are consistent with $\injindx=0.49\pm 0.1$ while further away, at a relative distance to the core of 0.5, the lobes are more consistent with a steeper $\injindx\approx0.65$.

\begin{figure*}[th]
	\centering
	\includegraphics[width=.9\columnwidth, trim={.cm 0.cm .cm .cm},clip]{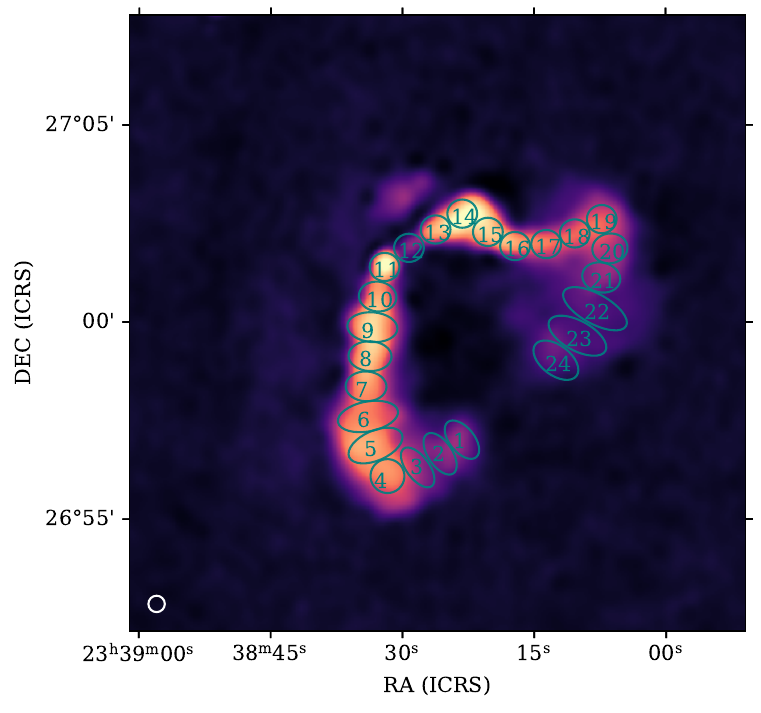 } 
	\includegraphics[width=.95\columnwidth, trim={.cm 0.cm .cm .cm},clip]{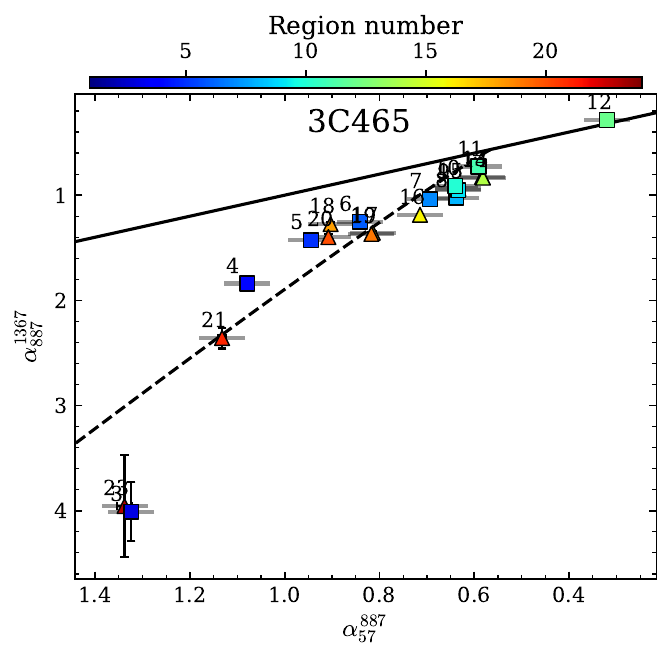 } 
	\caption{Left panel: LBA image of 3C~465 overlaid with the numbered regions that were used to obtain the colour-colour diagram in blue. Right panel: colour-colour diagram of the same source with corresponding numbering. The colour map represents the region numbering.}
	\label{fig:3c465-multi-component}
\end{figure*}

\subsection{Re-acceleration}
\label{sec:reacceleration}
Another interpretation for the spectral structure of 3C~465 is re-acceleration. Figure~\ref{fig:3c465-multi-component} shows the LBA image of 3C~465 overlaid with the regions used to obtain the colour-colour diagram. The deviation from the fitted ageing model between region 6 and 7 in the eastern lobe and region 16 and 17 in the western lobe corresponds to the steepening of the low frequency spectral index (see top middle panel of Figure~\ref{fig:spidx_ex}) from $\sim0.8$ to $\sim1.0$. It also corresponds to the sudden reduction of surface brightness in for instance the LBA image \citep{Boxelaarprep}. 

The right panel of Figure~\ref{fig:3c465-multi-component} shows a flattening of the high frequency part of the spectrum compared to ageing expectations around the previously mentioned points, while the low frequency spectral index is not significantly steeper or flatter. This suggests that re-acceleration is happening along the lobes. 3C~465 is known to have hotspots  (around region 11 and 14 in the figure) but the flattening happens beyond that point and is likely not related to the hotspot. A colour-colour diagram which includes very low frequencies is especially sensitive to re-acceleration. Since it mostly enhances high frequency flux density, re-acceleration can be made visible by a vertical offset upwards from the ageing expectation in this case. This is exactly what is visible in the figure.  

With current low frequency data it is uncertain what is physically happening in the lobes. VLA \citep{2020MNRAS.496..676B} and LoTSS images show some filamentary structure in the lobes at the flattening regions, but high-res olution ultra-low frequency data is missing to do a detailed analysis. After the upcoming upgrade of LOFAR 2.0, which coincides with a period of calmer ionospheric conditions, we will be able to obtain high-res olution images at 50~MHz.

\begin{table}[ht]

	\caption{\label{tab:results} Injection and spectral indices.}
	\centering
	\begin{tabular}{lccccccc}
		\hline\hline
		Object  & FR type\tablefootmark{a} & $\injindx$ & $\alpha_{58}^{144}$ & $\alpha_{58}^{144}$ \\
		&&&Hotspot&Lobe\\
		\hline
3C 192 &  II & $0.70 \pm 0.03$&0.88&$0.90-0.95$\\
3C 223 &  II & $0.46 \pm 0.06$&0.5&$0.65-0.75$\\
3C 236 &  II & $0.57 \pm 0.21$&0.8&$0.70-1.10$\\
3C 264 &  I & $0.51 \pm 0.26$&-&$0.60-0.80$\\
3C 284 &  II & $0.84 \pm 0.04$&0.79&$0.95-1.05$\\
3C 296 &  I & $0.43 \pm 0.02$&-&$0.55-0.60$\\
3C 31 &  I & $0.38 \pm 0.17$&-&$0.60-1.10$\\
3C 310 &  I & $0.60 \pm 0.05$&-&$0.80-0.95$\\
3C 326 &  II & $0.60 \pm 0.20$&0.49&$0.80-1.00$\\
3C 33 & II & $0.53 \pm 0.10$&&$0.60-1.10$\\
3C 382 & II & $0.69 \pm 0.04$&0.9&$0.85-0.95$\\
3C 442A &  II & $0.46 \pm 0.09$&0.75&$0.50-0.80$\\
3C 449 & I & $0.47 \pm 0.18$&-&$0.50-1.00$\\
3C 452 &  II & $0.68 \pm 0.02$&0.75&$0.85-1.00$\\
3C 457 & II & $0.78 \pm 0.09$&0.80&$0.90-1.40$\\
3C 46 & II & $0.84 \pm 0.02$&0.80&$0.85-0.95$\\
3C 465 & I & $0.49 \pm 0.10$&-&$0.65-1.15$\\
3C 76.1 & I & $0.67 \pm 0.13$&-&$0.85$\\
3C 98 &  II & $0.72 \pm 0.06$&0.61&$0.70-1.05$\\
4C 12.03 &  II & $0.68 \pm 0.07$&0.81&$0.95-1.05$\\
NGC 6109 &  I & $0.54 \pm 0.21$&-&$0.70-1.00$\\
NGC 7385 &  I & $0.45 \pm 0.10$&-&$0.80-1.50$\\
		\hline
	\end{tabular}
	\tablefoot{Fitted injection indices of the sub-sample of extended radio galaxies. Uncertainties are obtained from the RMSD mentioned in Section~\ref{sec:injidx}}
	\tablebib{(a)~\url{https://3crr.extragalactic.info/}}
\end{table}

\subsection{Steep low frequency spectral index} 

We observed a significant steepening in the measured injection index of both FR~I and FR~II galaxies when we considered $\alpha_{144}^{58}$ instead of $\alpha_{887}^{58}$. We also occasionally observed that the spectral index $\alpha_{144}^{58}$ between LBA and HBA is steeper than the spectral index $\alpha_{887}^{144}$ between HBA and ASKAP-low, contrary to what is expected from a typical ageing synchrotron spectrum. This could be explained by systematic overestimation of flux densities in LBA or underestimation in LoTSS. The flux density scales were thoroughly checked in Paper~I for the LBA data. There, we observed an under estimation of flux in LoTSS by a few percent, which could explain the different measurements in injection index. As mentioned in Section~\ref{sec:colour-colour}, we measured $\injindx$ from $\alpha^{887}_{58}$ to reduce the effect of possible flux scale offsets. 

Assuming there is no issue with the flux scale and measurements of all images, a physical explanation for the steepening is re-acceleration of electrons in the lobes enhancing emission around GHz frequencies. Alternatively, we could be observing a superposition of electron populations which mix at our resolution and result in a deviation from a simple ageing model or power-law.  

More high-res olution, multi-frequency observations are necessary to better understand particle acceleration and particle composition in the lobes of FR~Is and FR~IIs. A handful of 3C sources have been observed with LOFAR at 144~MHz and produced high fidelity sub-arcsecond images \citep{2023MNRAS.520.4427M}. Observation time has been allocated to observe 73 3CRR sources at high-res olution using JVLA in L-band and C-band. Additionally, observations with LOFAR LBA exist but have not yet been calibrated with long baselines due to the difficult ionospheric conditions at the time of observing. These new multi-frequency observations will enable us to understand the lobe structure and energetics in better detail.

\section{Conclusions}
\label{sec:conclusion}

This work has presented the first study of the well-resolved spectral structure of bright radio galaxies at ultra-low frequencies (<100~MHz) using spectral index maps of a sub-sample of extended sources in the 3CRR catalogue. This is the first spectral study of a medium-sized sample of 22 radio galaxies utilising the LOFAR LBA system at 58~MHz. These data have allowed us to study the low frequency spectral structure as well as to constrain the injection index $\injindx$ using additional high frequency data from LoTSS and ASKAP. Measurements of the injection index at the lowest frequencies have long been a missing piece to advance our knowledge of (re)acceleration mechanisms of electrons in the lobes of radio galaxies. We have produced low-frequency spectral index maps using complementary LoTSS DR3 data at 144~MHz and ASKAP-low and -mid data at 887~MHz and 1367~MHz respectively. We have additionally measured the injection index $\injindx$ for these sources from colour-colour diagrams. The spectral indices and injection indices are summarised in Table~\ref{tab:results}. Besides these data products, the key results from this paper are as follows:  

   \begin{enumerate}[(i)]
      \item the low-frequency (<150~MHz) spectral index in the hotspots of FR~II galaxies is steeper than the measured injection index of the same sources. The hotspot emission is thus not representative of the injected plasma which is ageing in the lobes. The hotspot emission is influenced by more complex processes, for which future higher resolution ultra-low frequency observations will be necessary.
      \\
      \item The FR~II injection index is consistently $\geq0.5$ and correlates with the total luminosity of the source. Injection indices for FR~IIs can reach up to 0.9.
      \\
      \item the low-frequency injection index of FR~Is is around 0.5 close to the core which corresponds to the measured injection index. The obtained injection indices for the FR~Is in our sample, despite their high uncertainty, suggest that the injection indices at ultra-low frequencies remain consistent with what has been found in earlier work.
      \\
      \item Approximately one in three sources has a spectral structure that is not well described by a JP ageing model, leading to high uncertainties on injection index. Due to the complex structure or projection effects seen in for instance 3C 76.1 or 3C 382, the question arises whether these sources are suitable for classic ageing analysis at all. X-shaped or fat lobes, in combination with projection effects and the limited resolution of the current data, can result in a superposition of emission from physically distinct regions along the line of sight, leading to spectral index structures that cannot be described by a simple ageing model.
      \\
      \item Mixing of cosmic ray electron populations of different ages has minimal influence on the estimation of the injection index when measured through a colour-colour diagram. This is true under the assumption that all electrons of which emission is observed originate from the same injection source.
      Contrary to the injection index, the local spectral index is of course susceptible to mixing effects. 
   \end{enumerate}

\begin{acknowledgements}
JMB and FdG acknowledge the support of the ERC Consolidator Grant ULU 101086378. MJH thanks the UK STFC for support [ST/Y001249/1].
EDR acknowledges support by the Deutsche Forschungsgemeinschaft (DFG).
LOFAR is the Low Frequency Array designed and constructed by ASTRON. It has observing, data processing, and data storage facilities in several countries, which are owned by various parties (each with their own funding sources), and which are collectively operated by the LOFAR ERIC under a joint scientific policy. The LOFAR resources have benefited from the following recent major funding sources: CNRS-INSU, Observatoire de Paris and Université d'Orléans, France; BMFTR, MKW-NRW, MPG, Germany; Science Foundation Ireland (SFI), Department of Business, Enterprise and Innovation (DBEI), Ireland; NWO, The Netherlands; The Science and Technology Facilities Council, UK; Ministry of Science and Higher Education, Poland; The Istituto Nazionale di Astrofisica (INAF), Italy.

This research made use of the Dutch national e-infrastructure with support of the SURF Cooperative (e-infra 180169) and the LOFAR e-infra group. The Jülich LOFAR Long Term Archive and the German LOFAR network are both coordinated and operated by the Jülich Supercomputing Centre (JSC), and computing resources on the supercomputer JUWELS at JSC were provided by the Gauss Centre for Supercomputing e.V. (grant CHTB00) through the John von Neumann Institute for Computing (NIC).

This research made use of the University of Hertfordshire high-performance computing facility and the LOFAR-UK computing facility located at the University of Hertfordshire and supported by STFC [ST/P000096/1], and of the Italian LOFAR-IT computing infrastructure supported and operated by INAF, including the resources within the PLEIADI special 'LOFAR' project by USC-C of INAF, and by the Physics Department of Turin university (under an agreement with Consorzio Interuniversitario per la Fisica Spaziale) at the C3S Supercomputing Centre, Italy.

This research is part of the project LOFAR Data Valorization (LDV) [project numbers 2020.031, 2022.033, and 2024.047] of the research programme Computing Time on National Computer Facilities using SPIDER that is (co-)funded by the Dutch Research Council (NWO), hosted by SURF through the call for proposals of Computing Time on National Computer Facilities.

This scientific work uses data obtained from Inyarrimanha Ilgari Bundara / the Murchison Radio-astronomy Observatory. We acknowledge the Wajarri Yamaji People as the Traditional Owners and native title holders of the Observatory site. CSIRO’s ASKAP radio telescope is part of the Australia Telescope National Facility (https://ror.org/05qajvd42). Operation of ASKAP is funded by the Australian Government with support from the National Collaborative Research Infrastructure Strategy. ASKAP uses the resources of the Pawsey Supercomputing Research Centre. Establishment of ASKAP, Inyarrimanha Ilgari Bundara, the CSIRO Murchison Radio-astronomy Observatory and the Pawsey Supercomputing Research Centre are initiatives of the Australian Government, with support from the Government of Western Australia and the Science and Industry Endowment Fund. This paper includes archived data obtained through the CSIRO ASKAP Science Data Archive, CASDA (\url{https://data.csiro.au}).

This research has made use of the NASA/IPAC Extragalactic Database (NED), which is operated by the Jet Propulsion Laboratory, California Institute of Technology, under contract with the National Aeronautics and Space Administration. 
This research has made use of the VizieR catalogue access tool, CDS,
Strasbourg, France (DOI : 10.26093/cds/vizier). The original description 
of the VizieR service was published in \citet{vizier2000}.
This research made use of Astropy (\url{http://www.astropy.org}), a community-developed core Python package for Astronomy \citep{2013A&A...558A..33A, 2018AJ....156..123A}
\end{acknowledgements}

\bibliography{aa59503-26}

\onecolumn
\begin{appendix} 
\section{Low- and high-frequency spectral index maps }
\label{app:spidx_maps_all}

\begin{figure}[H]
\centering
\includegraphics[width=0.49\linewidth, trim={0.cm 0.cm 0.cm 0.cm},clip]{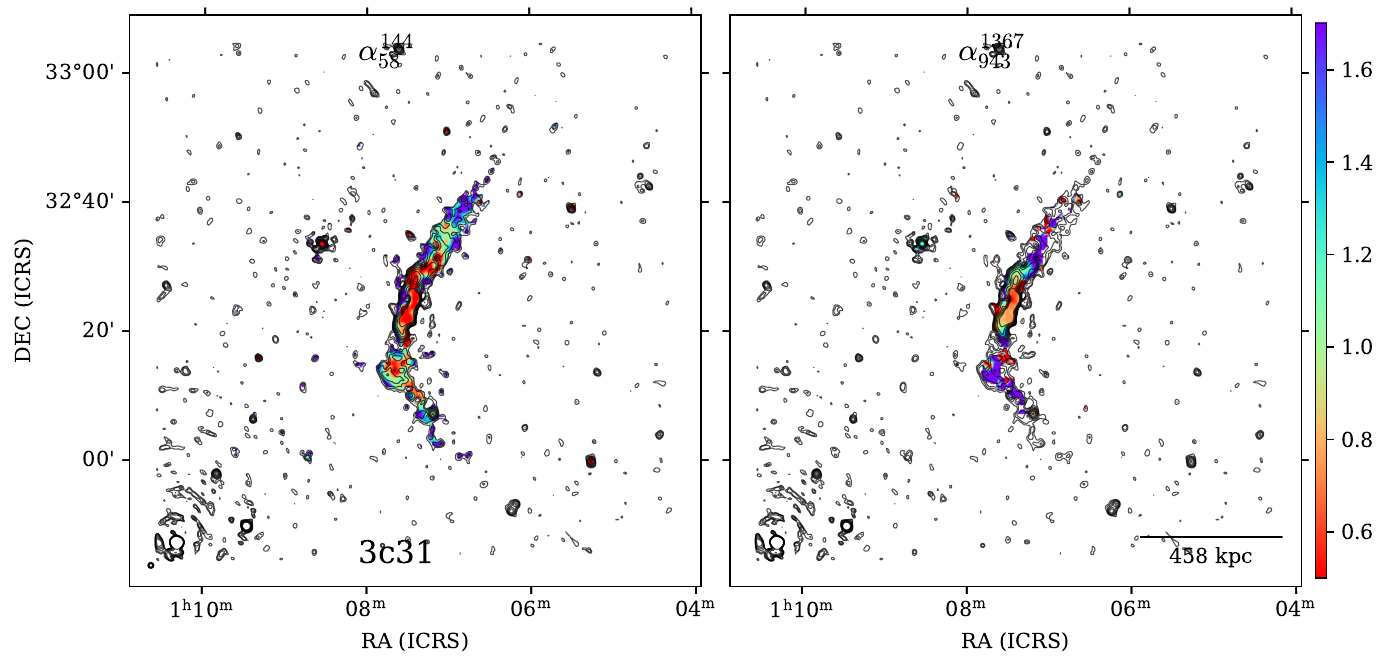}
\includegraphics[width=0.49\linewidth, trim={0.cm 0.cm 0.cm 0.cm},clip]{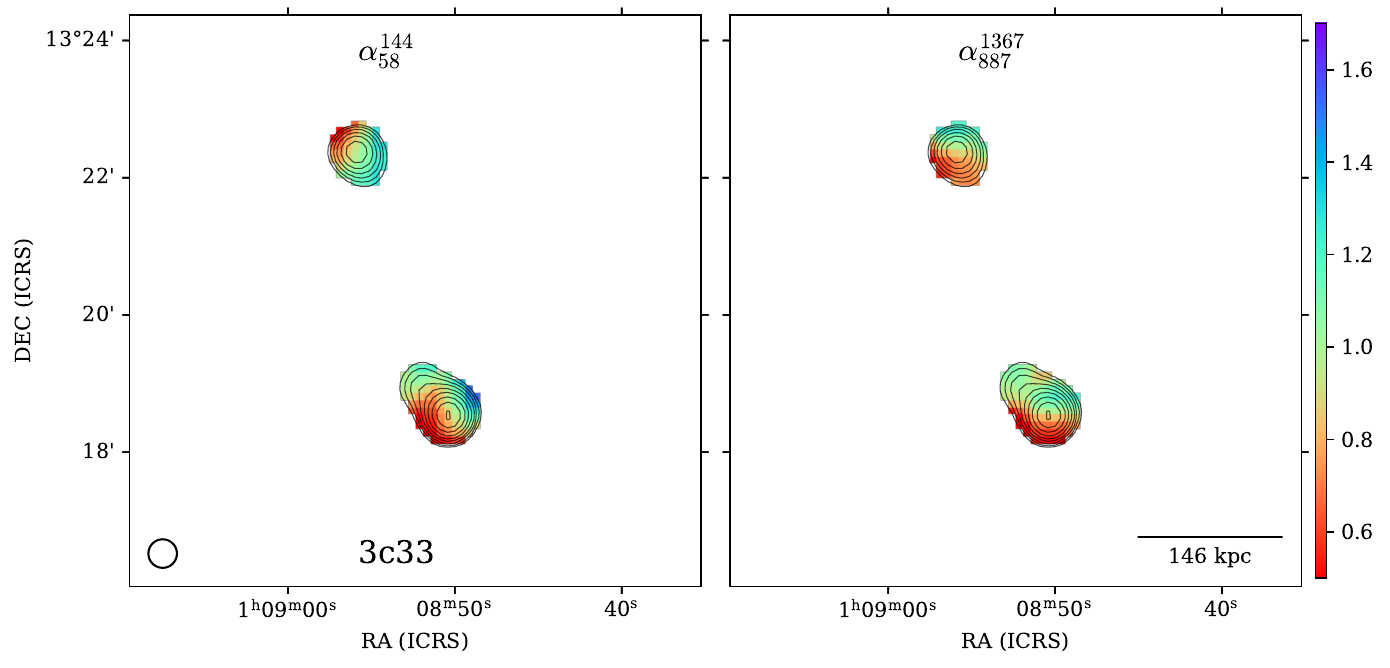}
\includegraphics[width=0.49\linewidth, trim={0.cm 0.cm 0.cm 0.cm},clip]{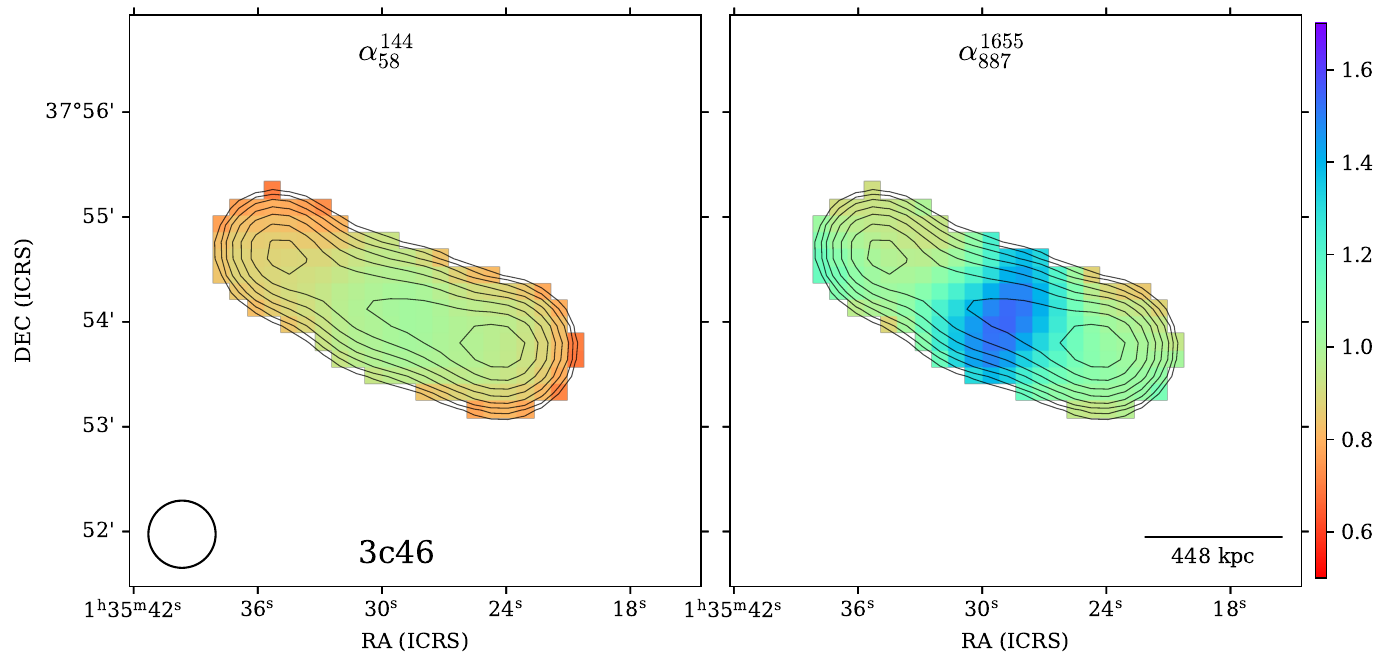}
\includegraphics[width=0.49\linewidth, trim={0.cm 0.cm 0.cm 0.cm},clip]{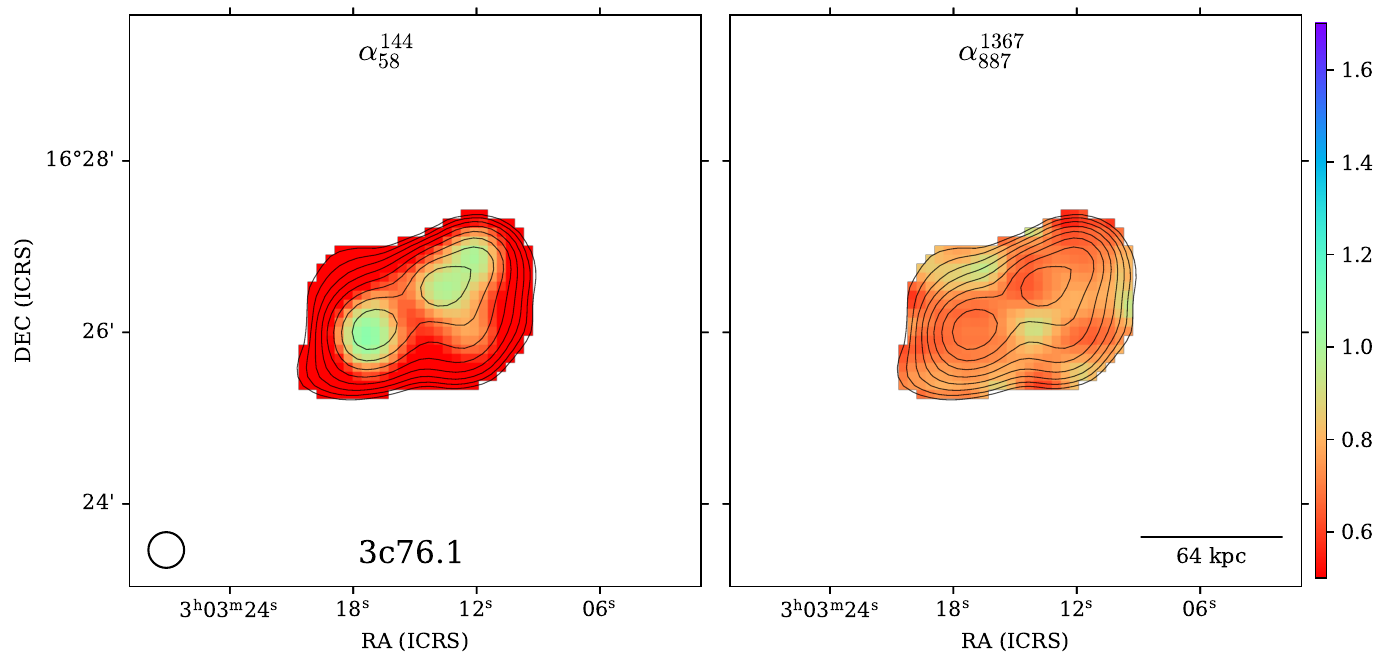}
\includegraphics[width=0.49\linewidth, trim={0.cm 0.cm 0.cm 0.cm},clip]{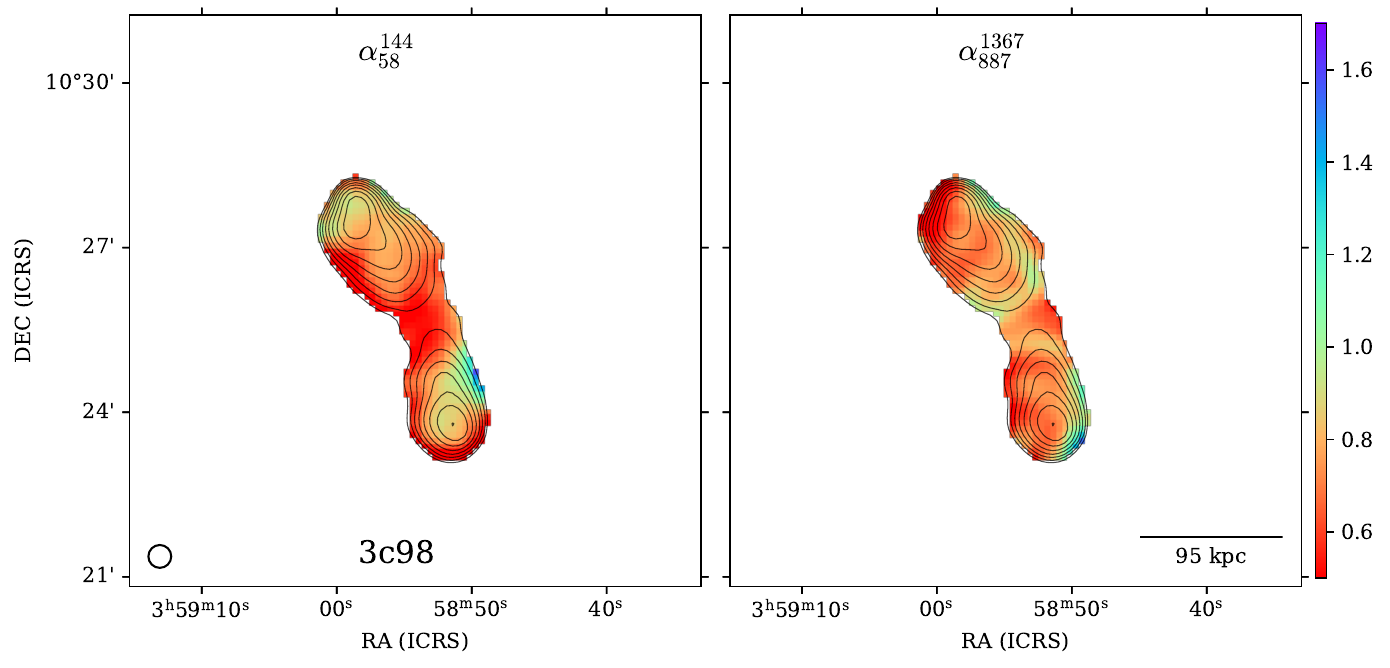}
\includegraphics[width=0.49\linewidth, trim={0.cm 0.cm 0.cm 0.cm},clip]{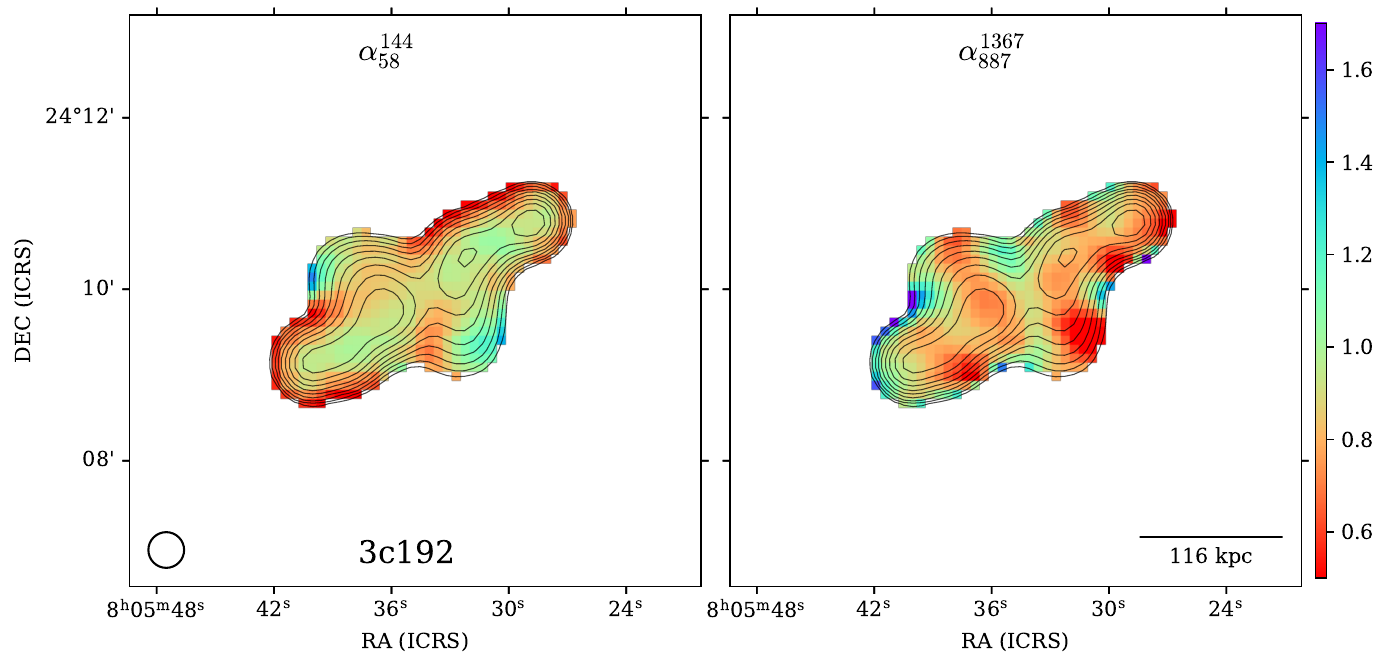}
\includegraphics[width=0.49\linewidth, trim={0.cm 0.cm 0.cm 0.cm},clip]{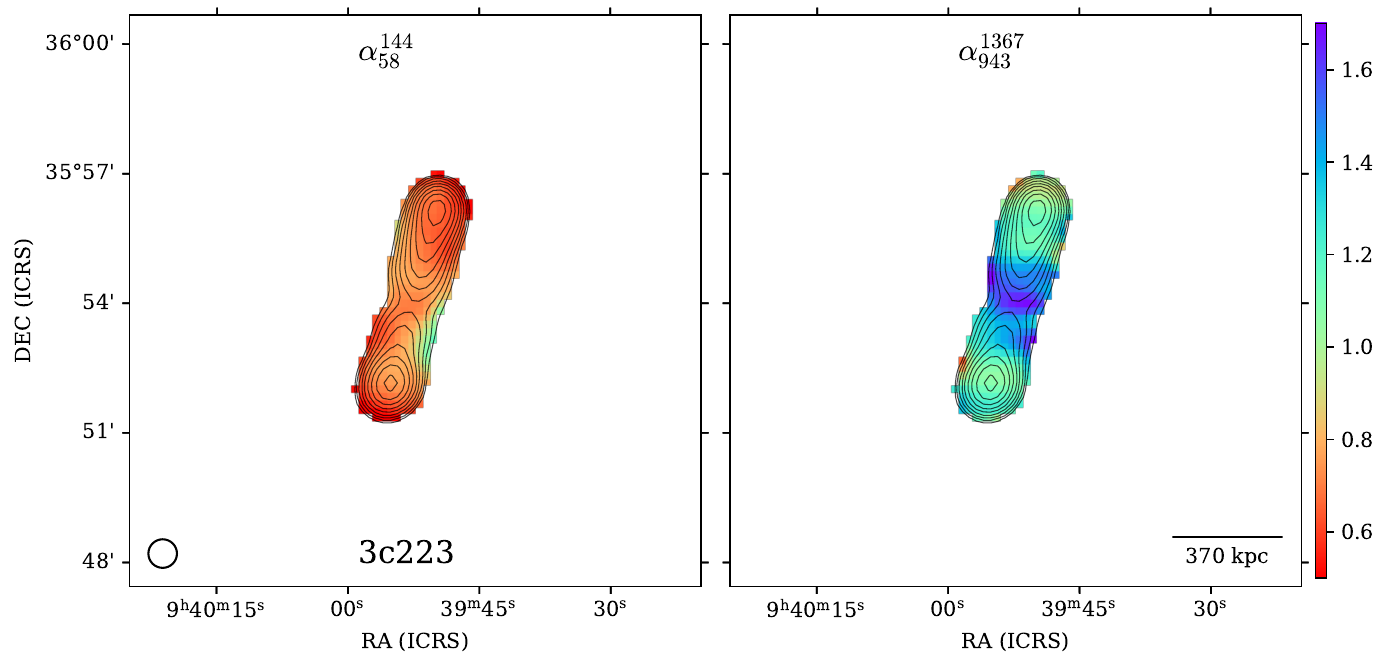}
\includegraphics[width=0.49\linewidth, trim={0.cm 0.cm 0.cm 0.cm},clip]{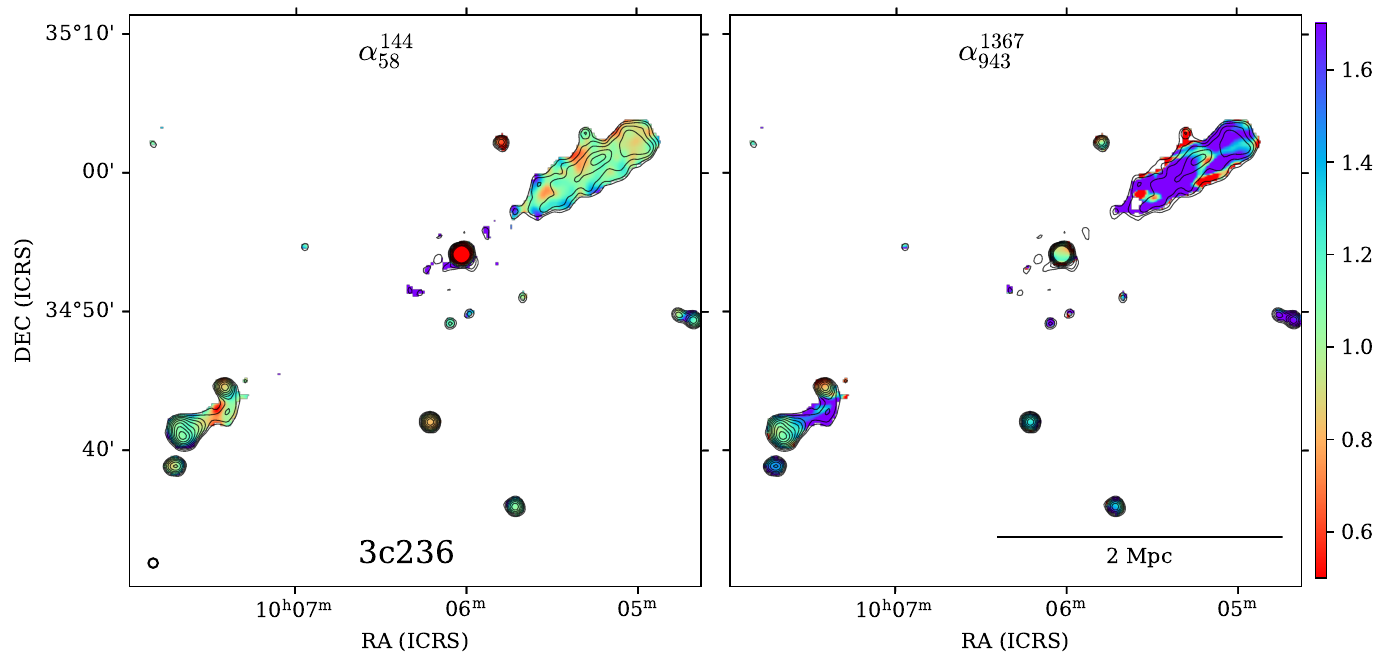}
\includegraphics[width=0.49\linewidth, trim={0.cm 0.cm 0.cm 0.cm},clip]{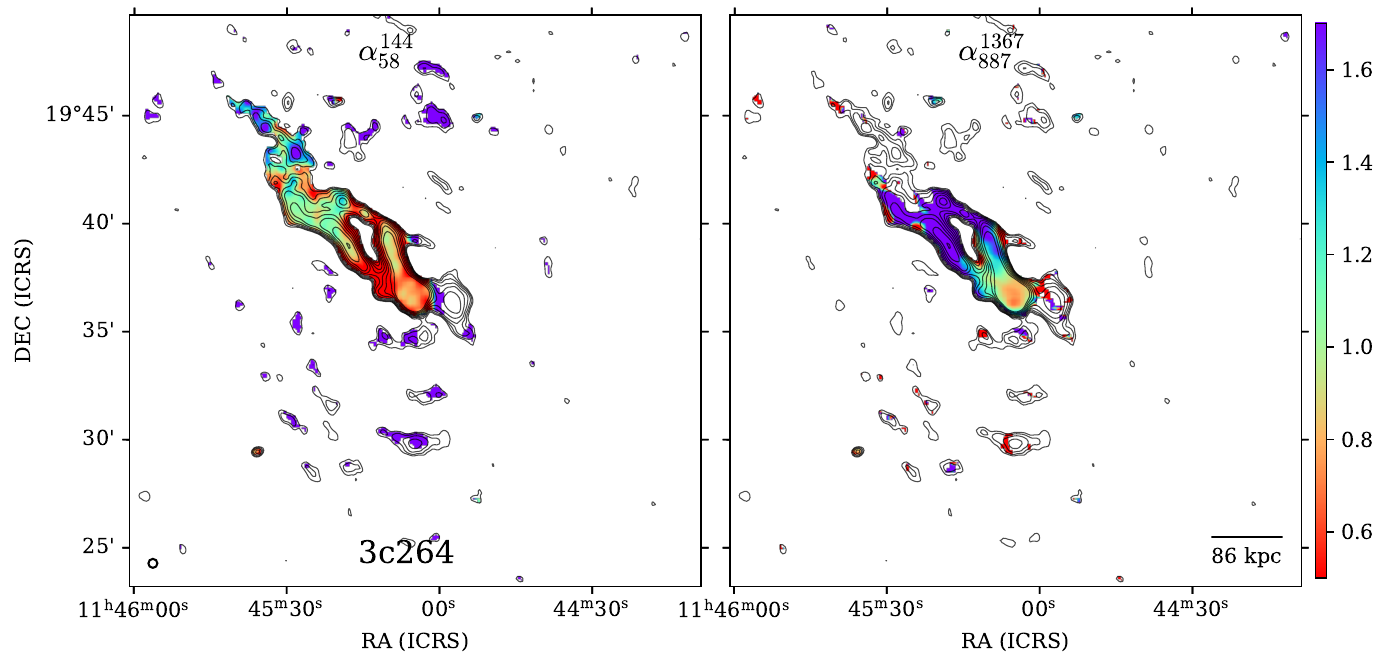}
\includegraphics[width=0.49\linewidth, trim={0.cm 0.cm 0.cm 0.cm},clip]{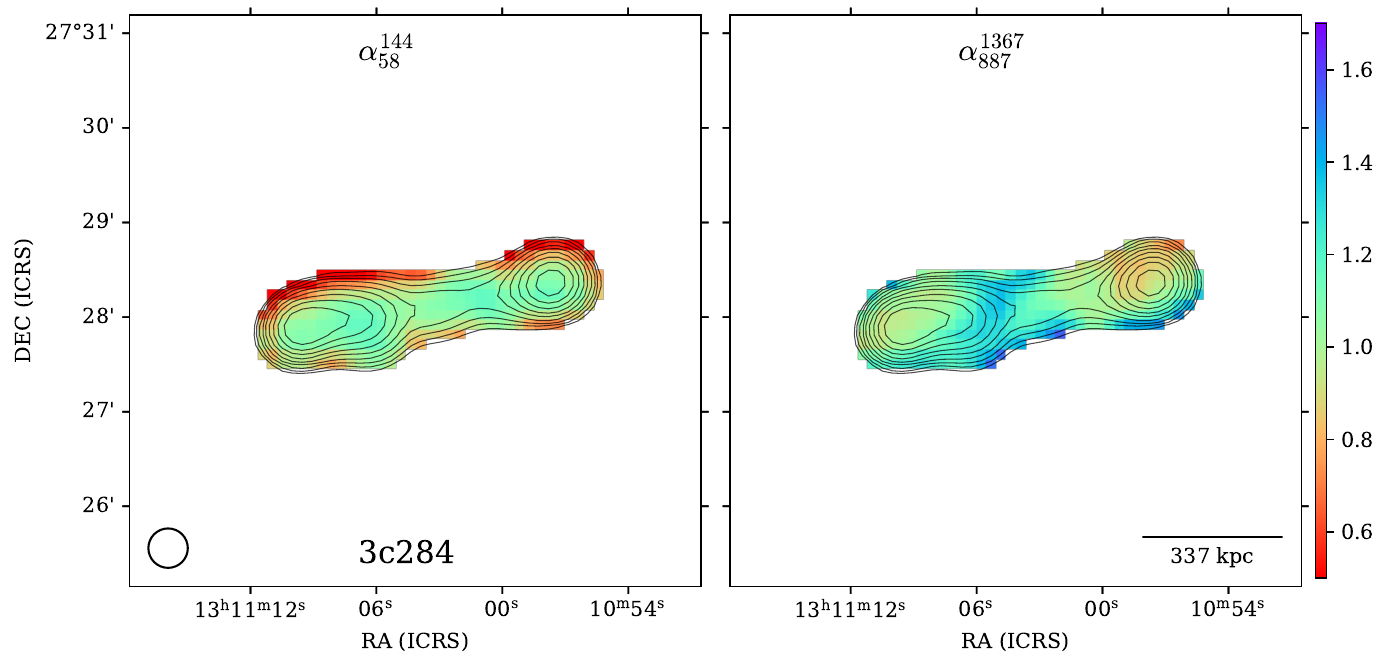}

\caption{Radio maps of 3C sources in our sample. Same as Figure~\ref{fig:spidx_ex}. The contours are drawn at $[9, 50, 100, 200]\times \sigma_{\text{rms}}$ by default, $[5, 10, 50, 100, 200]\times \sigma_{\text{rms}}$ for extended sources and $[3, 5, 10, 50, 100, 200]\times \sigma_{\text{rms}}$ for very extended sources.}
\end{figure}
\setcounter{figure}{0} 
\newpage

\begin{figure}[H]
\centering
\includegraphics[width=0.49\linewidth, trim={0.cm 0.cm 0.cm 0.cm},clip]{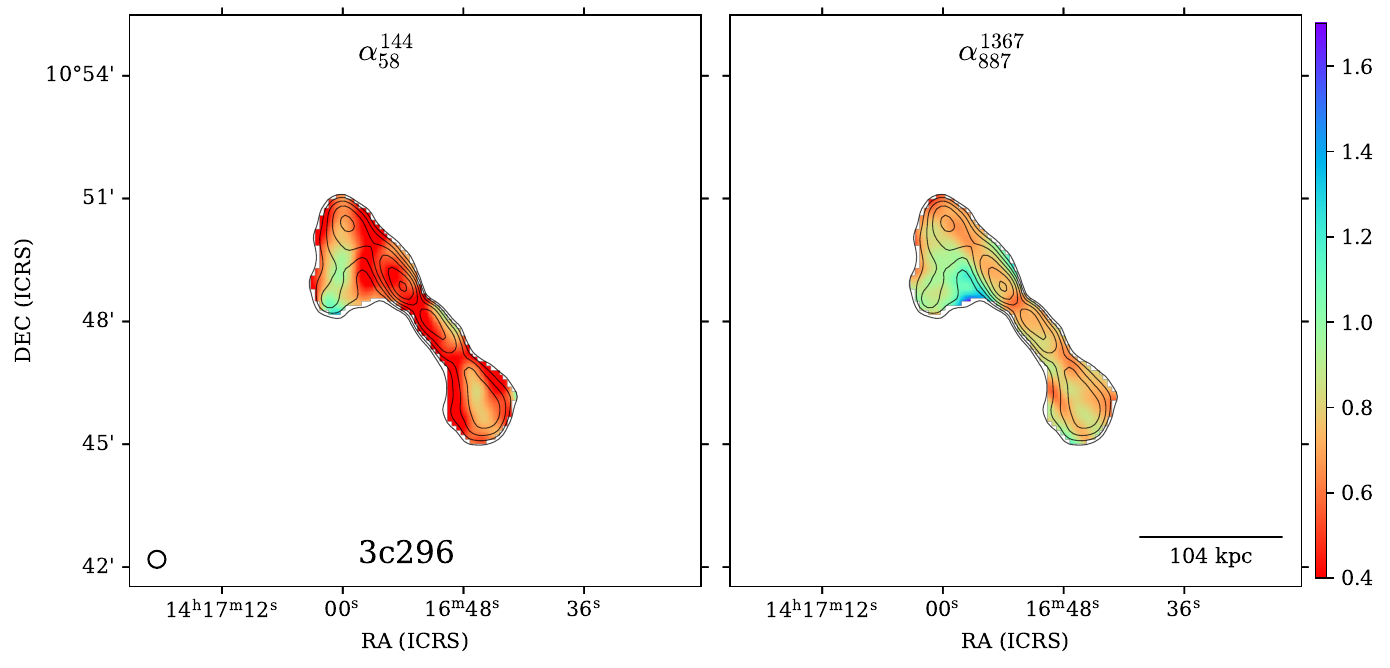}
\includegraphics[width=0.49\linewidth, trim={0.cm 0.cm 0.cm 0.cm},clip]{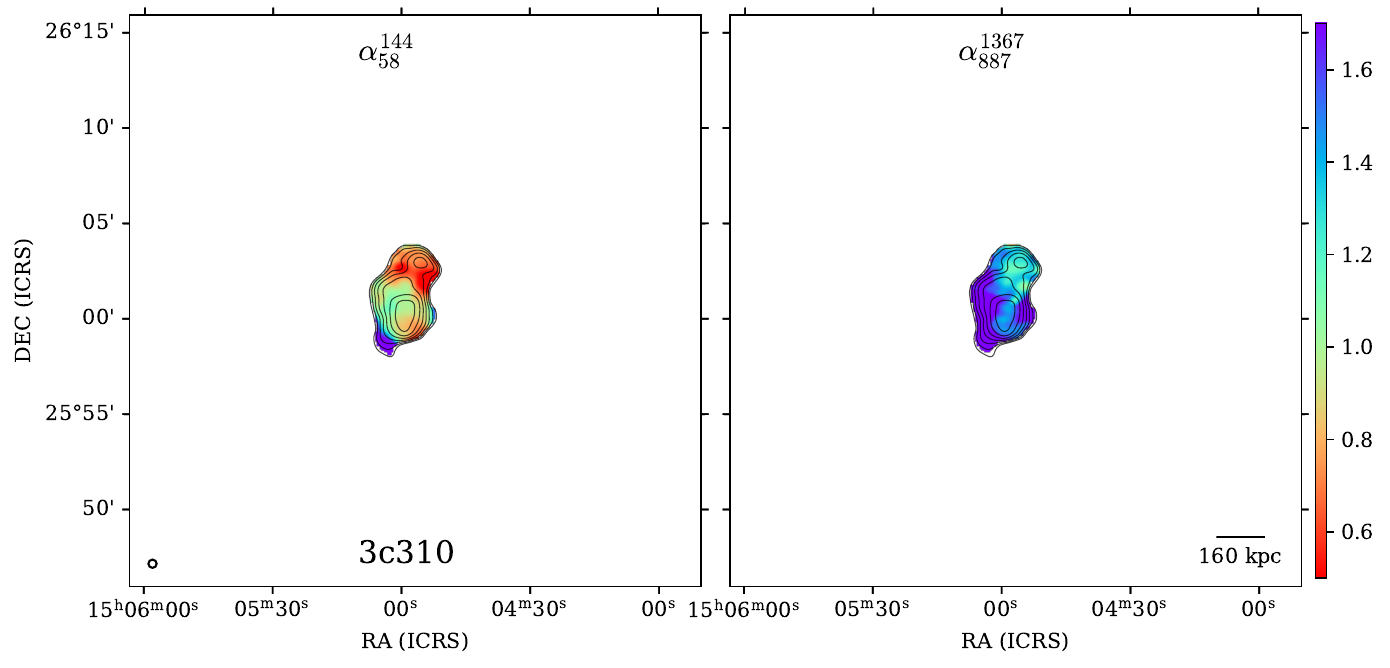}
\includegraphics[width=0.49\linewidth, trim={0.cm 0.cm 0.cm 0.cm},clip]{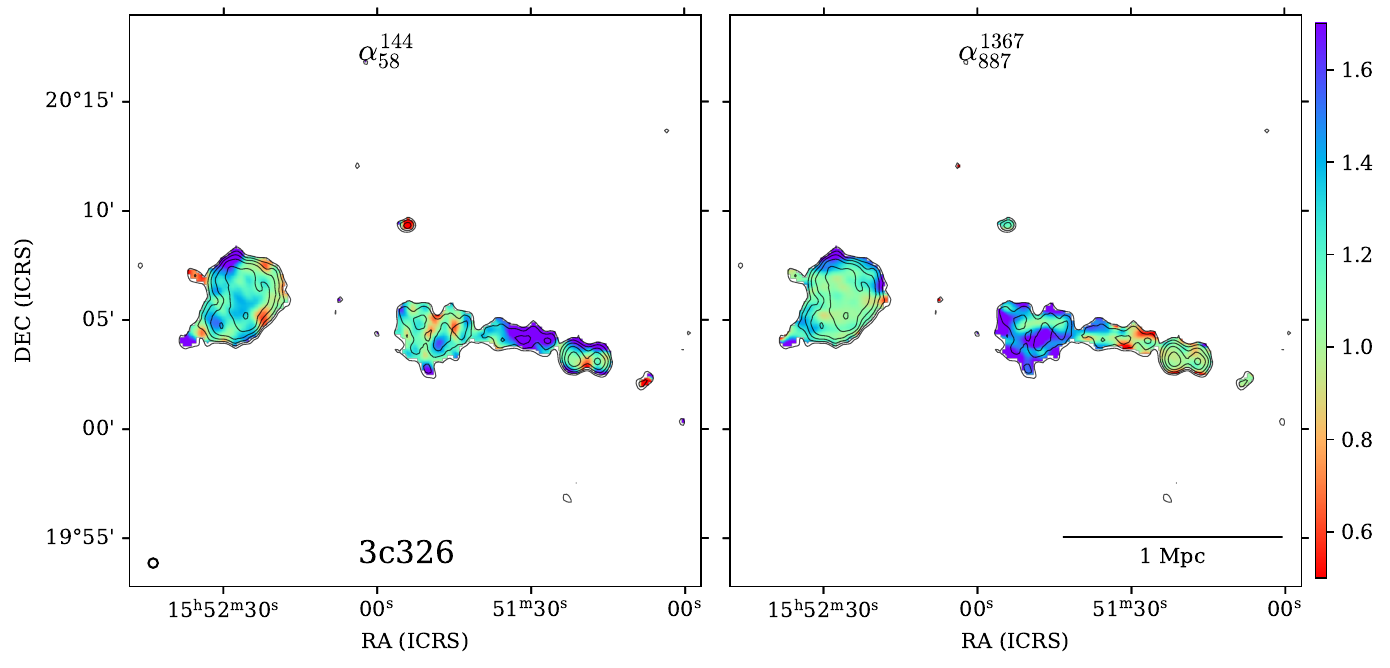}
\includegraphics[width=0.49\linewidth, trim={0.cm 0.cm 0.cm 0.cm},clip]{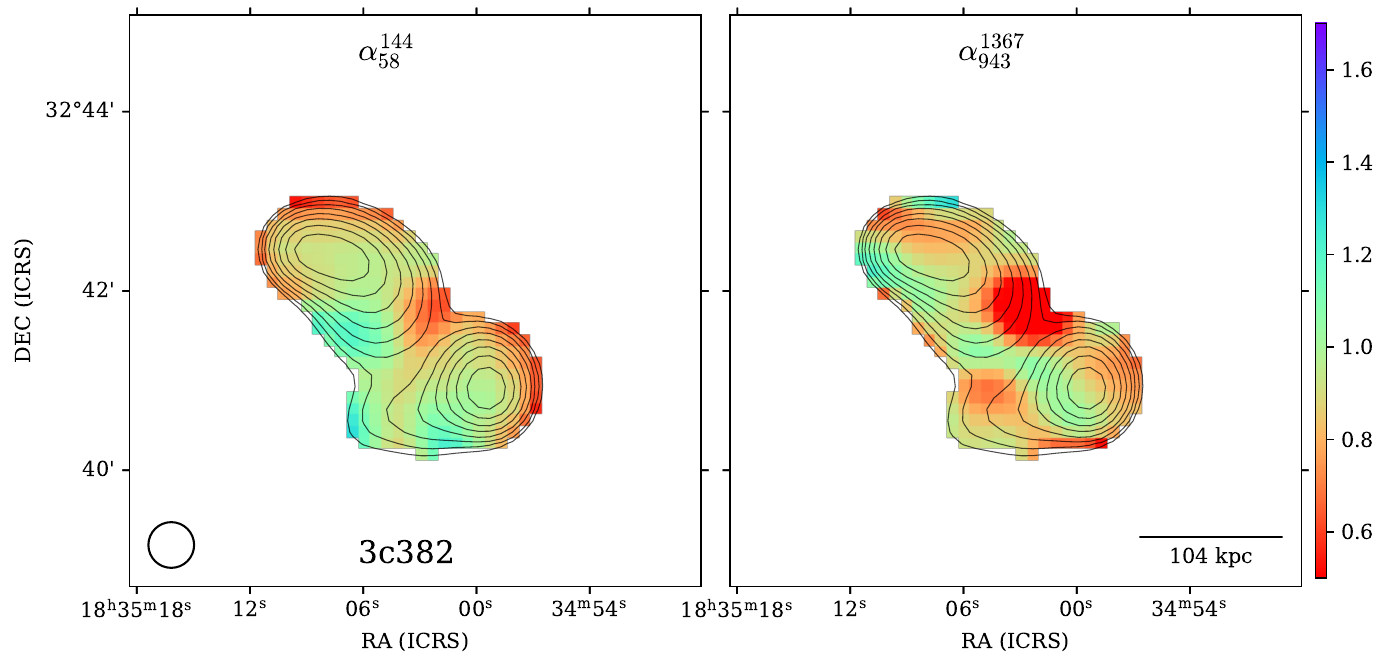}
\includegraphics[width=0.49\linewidth, trim={0.cm 0.cm 0.cm 0.cm},clip]{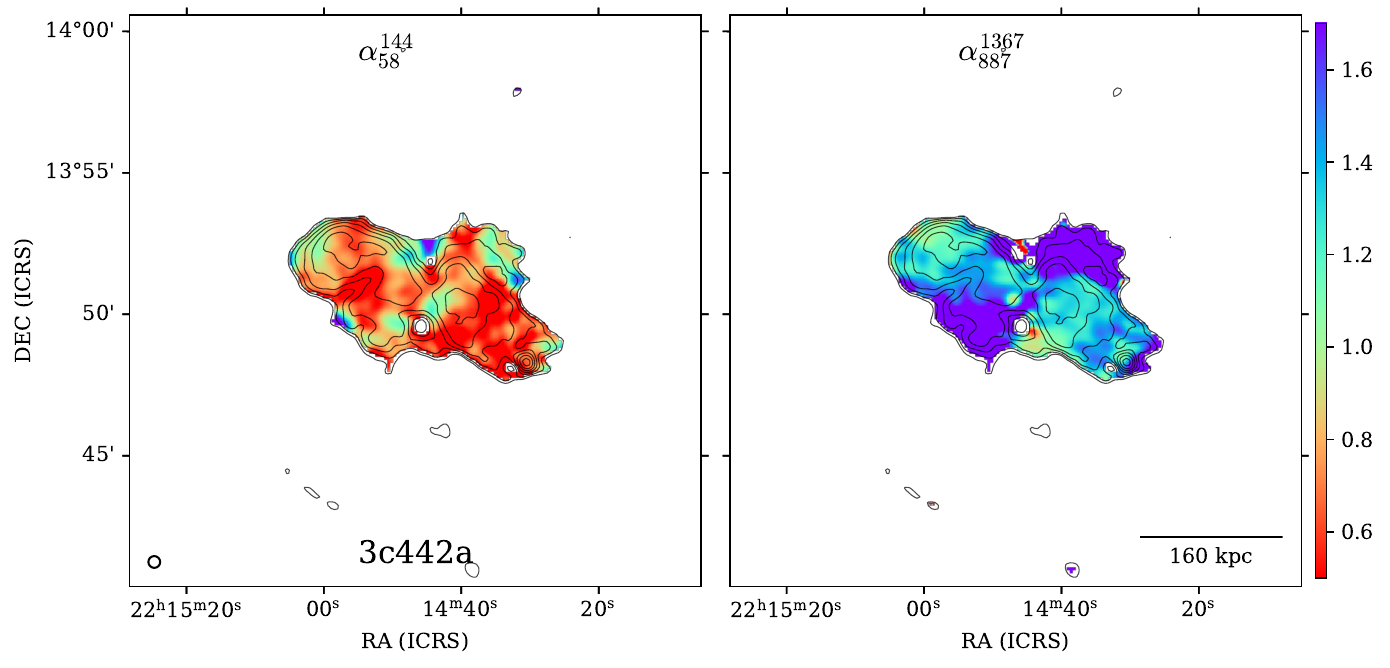}
\includegraphics[width=0.49\linewidth, trim={0.cm 0.cm 0.cm 0.cm},clip]{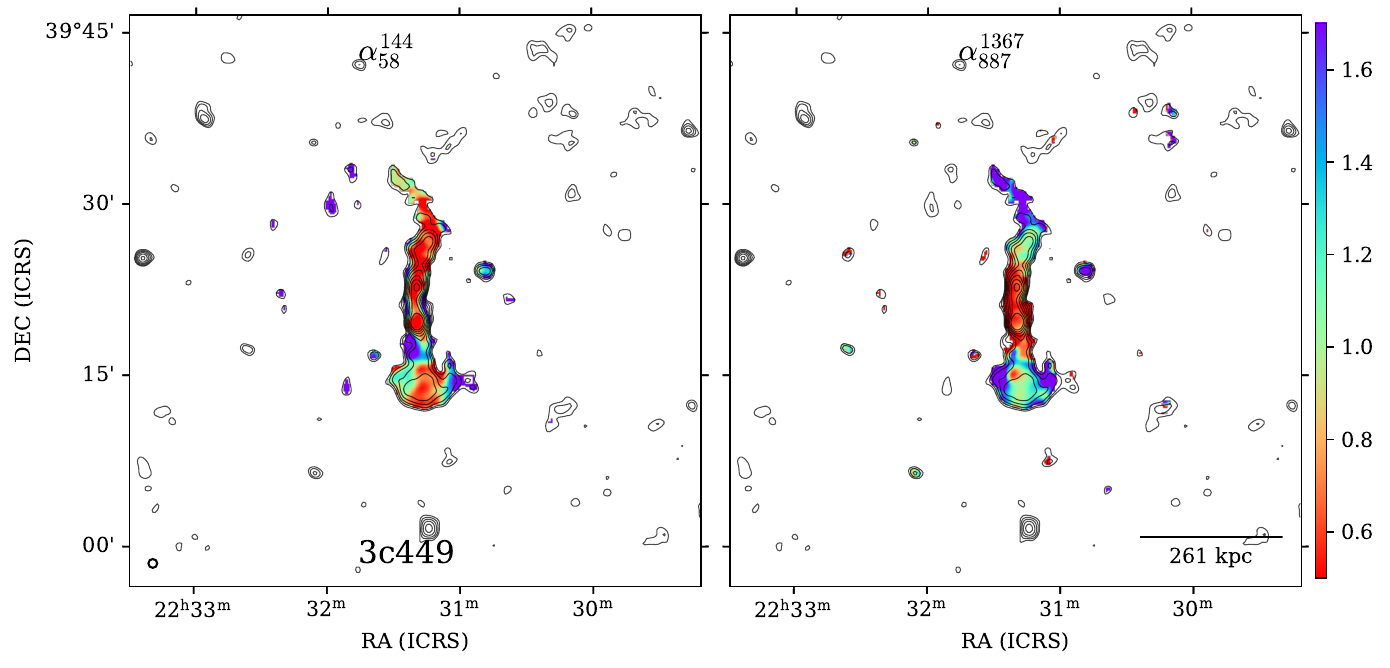}
\includegraphics[width=0.49\linewidth, trim={0.cm 0.cm 0.cm 0.cm},clip]{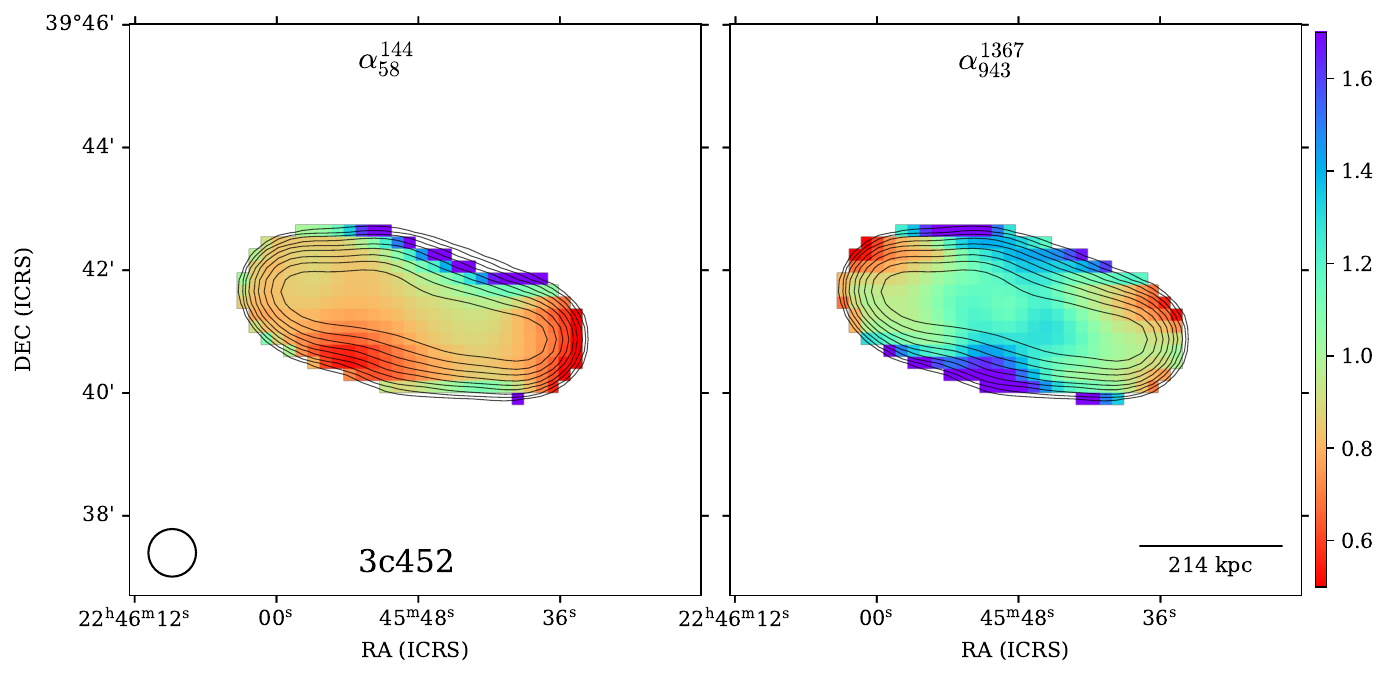}
\includegraphics[width=0.49\linewidth, trim={0.cm 0.cm 0.cm 0.cm},clip]{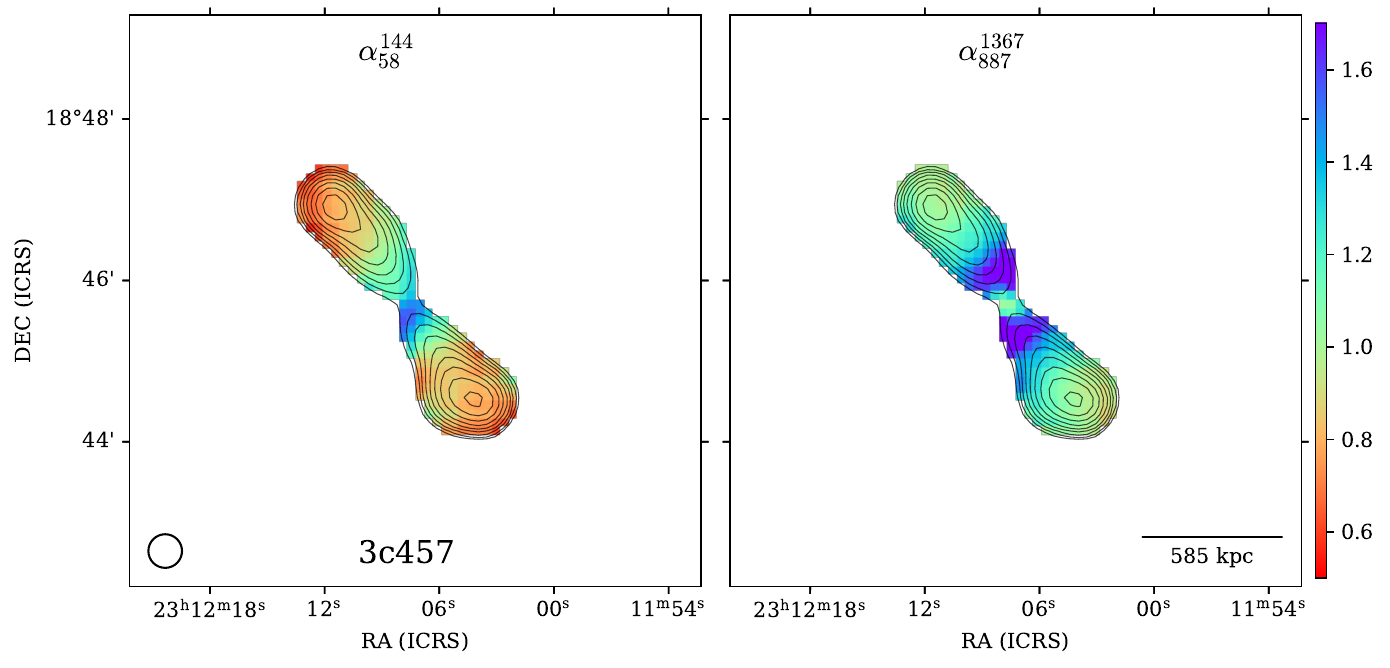}
\includegraphics[width=0.49\linewidth, trim={0.cm 0.cm 0.cm 0.cm},clip]{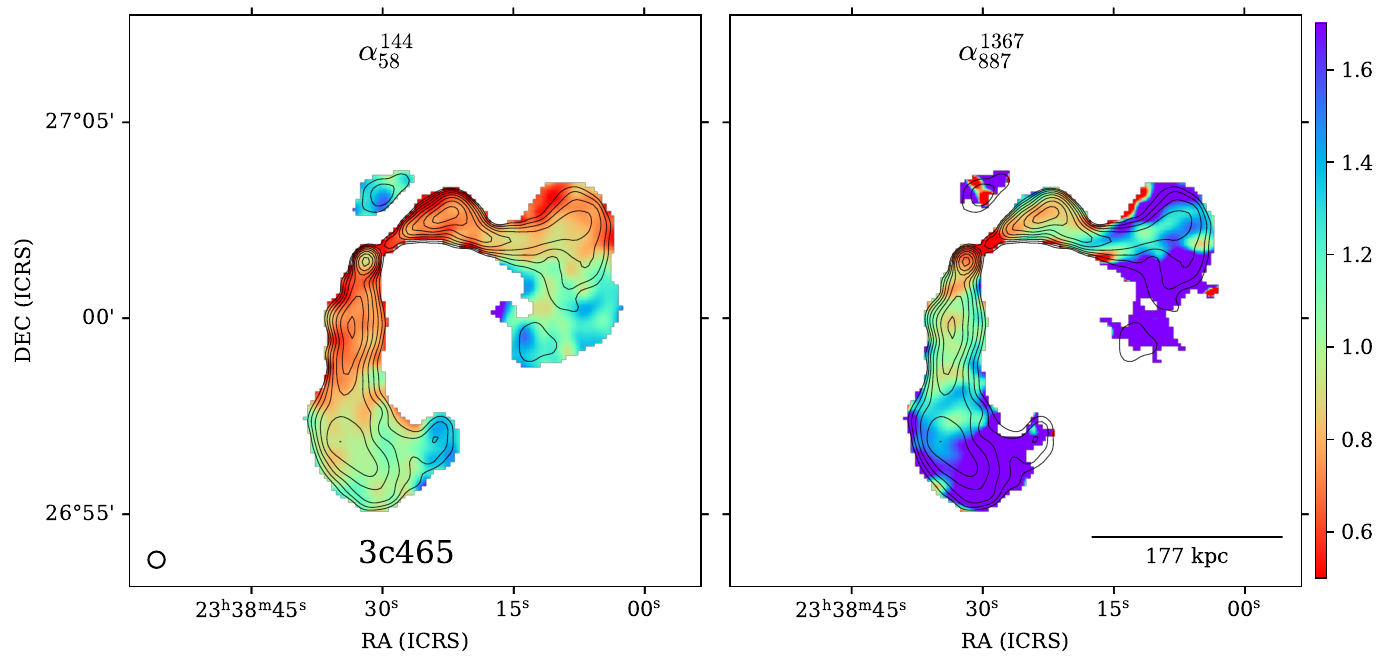}
\includegraphics[width=0.49\linewidth, trim={0.cm 0.cm 0.cm 0.cm},clip]{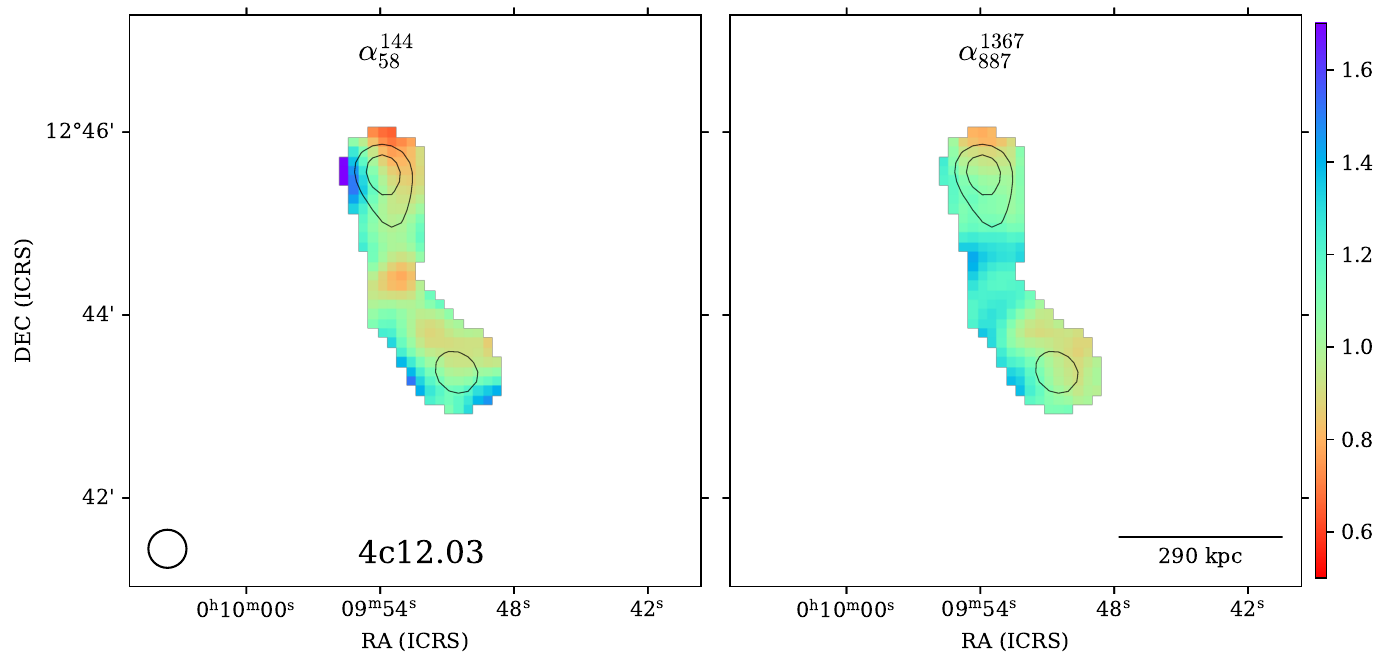}
\caption{Continued}
\end{figure}
\setcounter{figure}{0} 

\newpage
\begin{figure}[ht]
\includegraphics[width=0.49\linewidth, trim={0.cm 0.cm 0.cm 0.cm},clip]{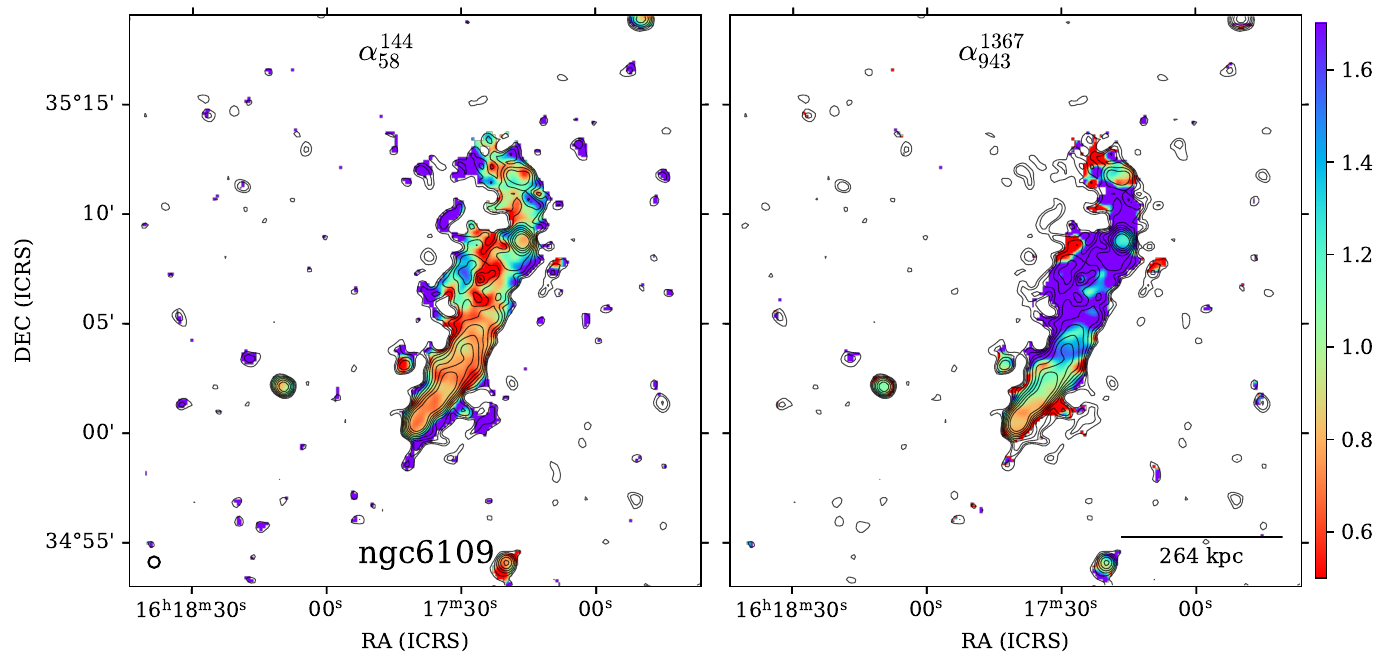}
\includegraphics[width=0.49\linewidth, trim={0.cm 0.cm 0.cm 0.cm},clip]{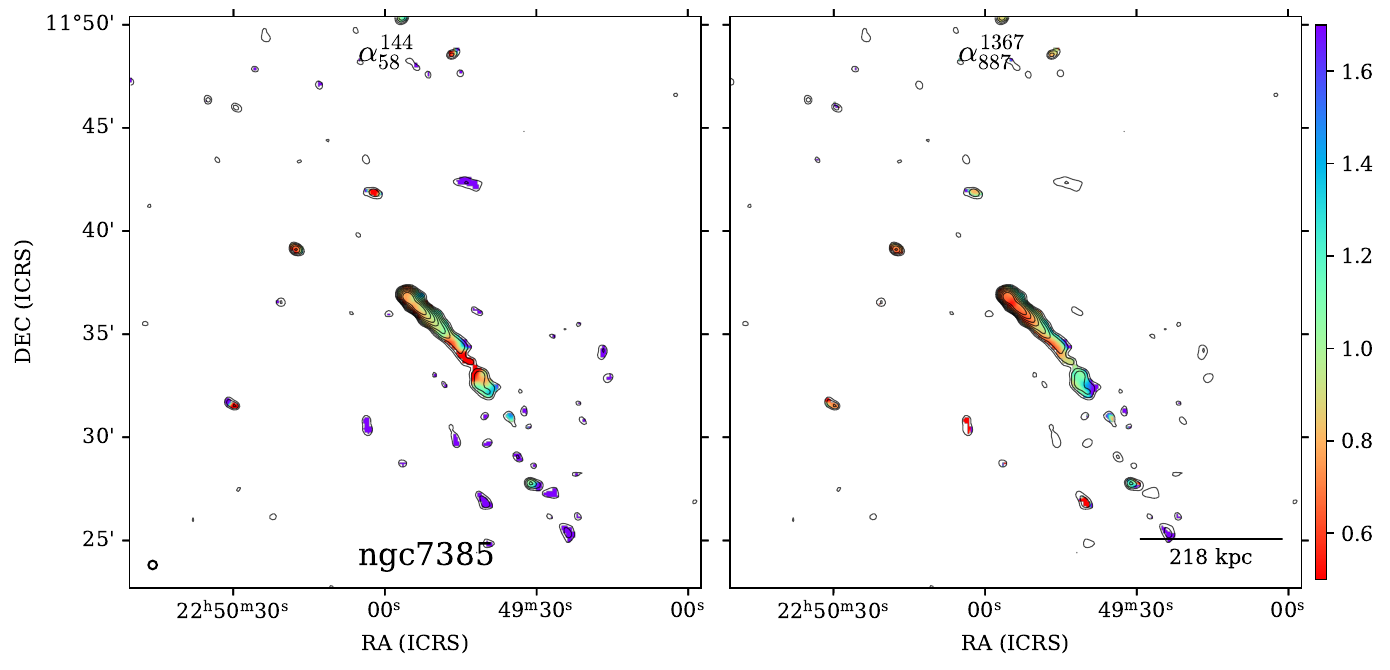}
\caption{Continued}
\end{figure}
\setcounter{figure}{0} 

\section{Colour-colour diagrams}
\label{app:colour-colour-diagram}
\begin{figure}[ht]
\centering
\includegraphics[width=0.27\linewidth, trim={0.cm 0.cm 0.cm 0.cm},clip]{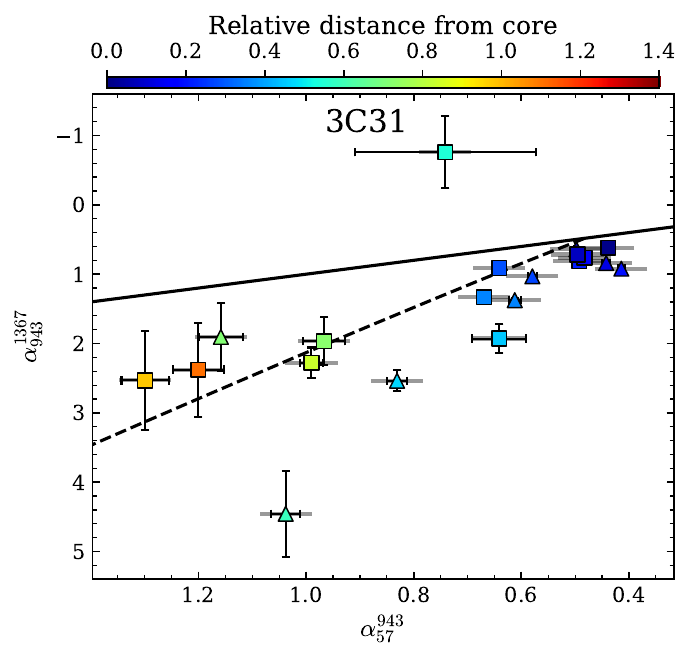}
\includegraphics[width=0.27\linewidth, trim={0.cm 0.cm 0.cm 0.cm},clip]{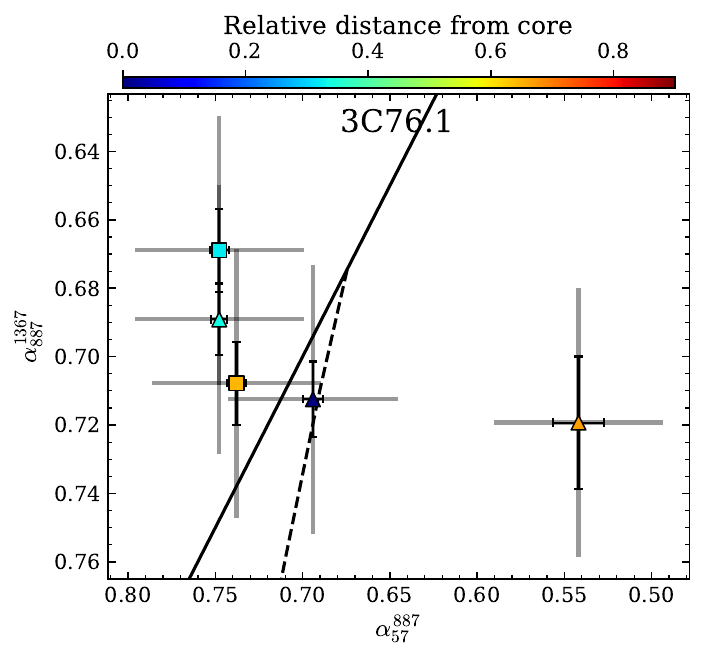}
\includegraphics[width=0.27\linewidth, trim={0.cm 0.cm 0.cm 0.cm},clip]{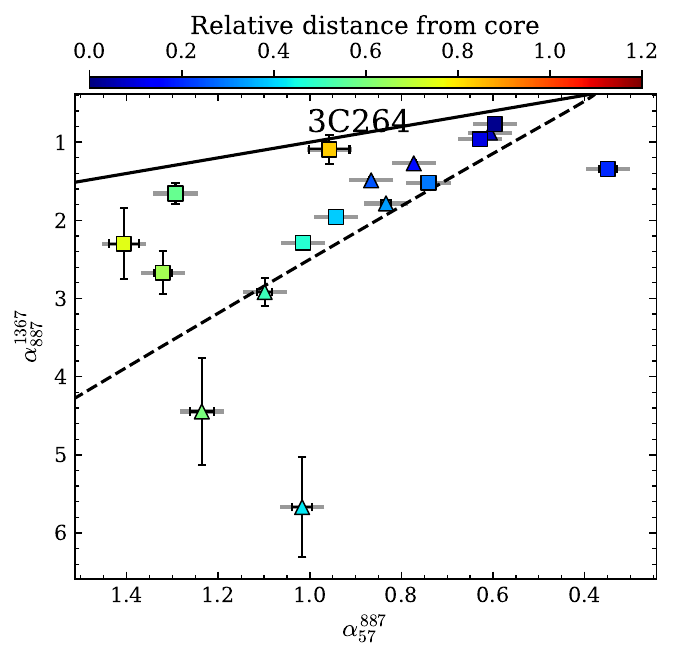}
\includegraphics[width=0.27\linewidth, trim={0.cm 0.cm 0.cm 0.cm},clip]{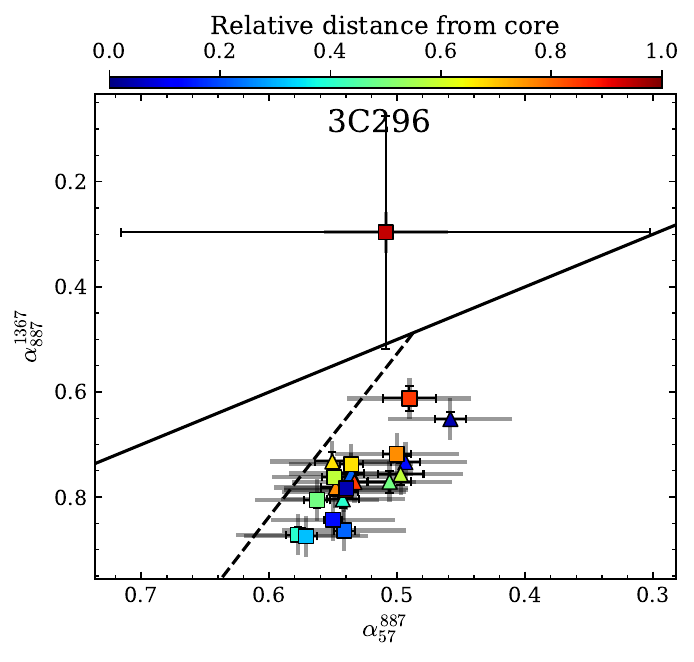}
\includegraphics[width=0.27\linewidth, trim={0.cm 0.cm 0.cm 0.cm},clip]{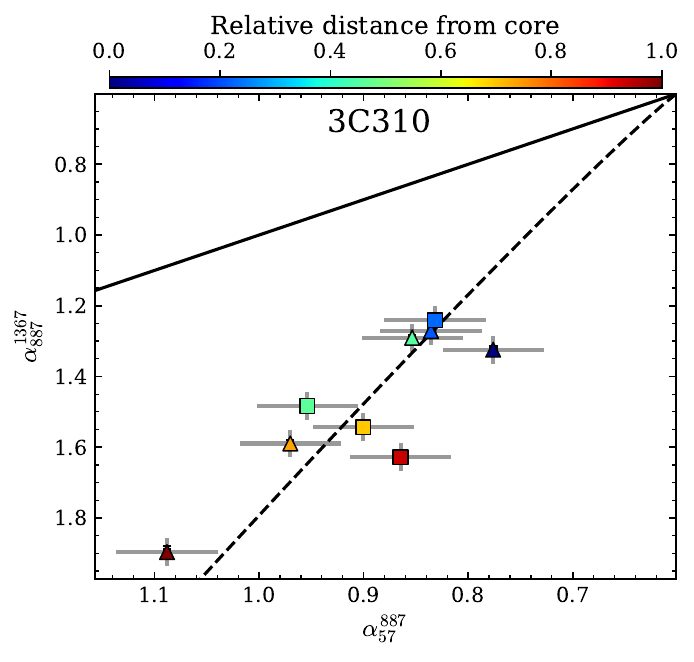}
\includegraphics[width=0.27\linewidth, trim={0.cm 0.cm 0.cm 0.cm},clip]{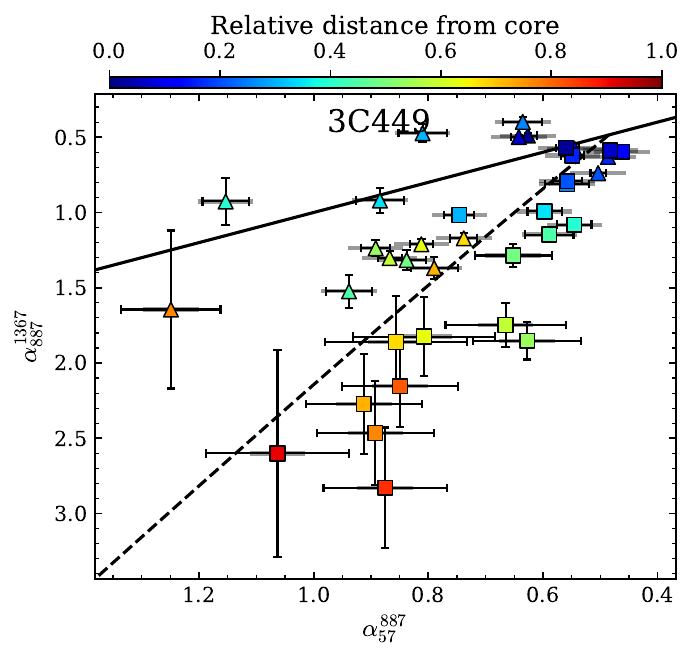}
\includegraphics[width=0.27\linewidth, trim={0.cm 0.cm 0.cm 0.cm},clip]{Assets/color_color/color_color_3c465___0,2_,_2,3__.pdf}
\includegraphics[width=0.27\linewidth, trim={0.cm 0.cm 0.cm 0.cm},clip]{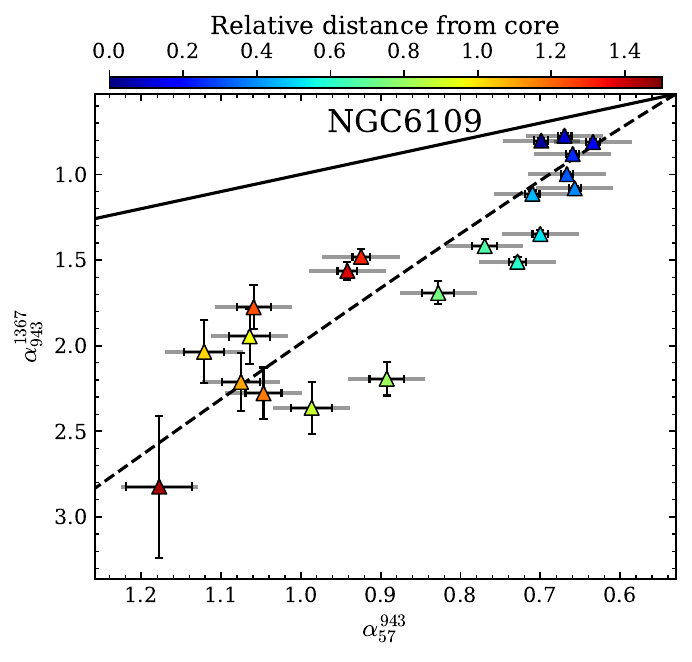}
\includegraphics[width=0.27\linewidth, trim={0.cm 0.cm 0.cm 0.cm},clip]{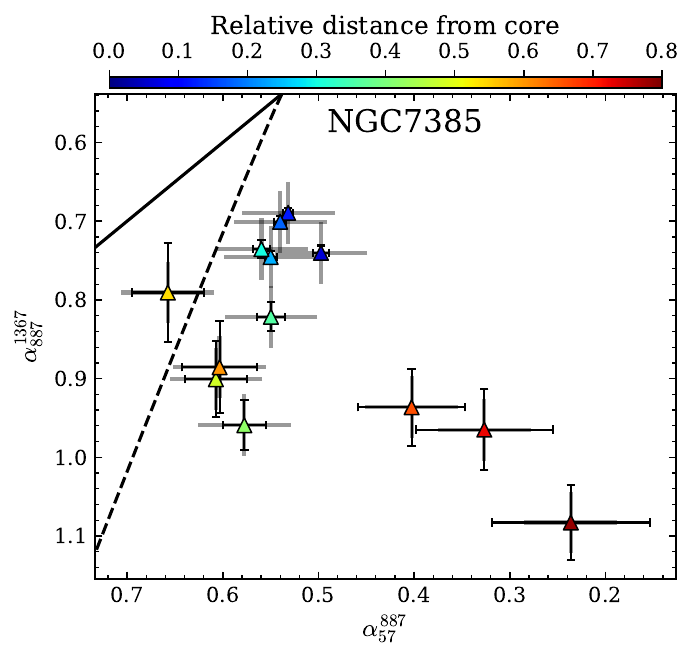}
\caption{Colour-colour diagrams of FR~I galaxies in Table~\ref{tab:the_subsample} in the $\alpha_{57}^{144}-\alpha_{887}^{1637}$ (or $\alpha_{57}^{144}-\alpha_{943}^{1367}$) plane. The colour scale denotes the distance from the core relative to the approximate size in Table~\ref{tab:the_subsample}. The solid grey error bar denotes the possible systematic shift of data points due to calibration errors and the thin black error bars denote the rms flux uncertainty. The solid line denotes a pure power-law and the dashed line a JP model.
The two lobes are differentiated by either showing a square or triangle. Regions containing the core are excluded from the figure. }
\label{fig:colour-colour-all-fri}
\end{figure}

\newpage
\begin{figure}[h]
\centering
\includegraphics[width=0.27\linewidth, trim={0.cm 0.cm 0.cm 0.cm},clip]{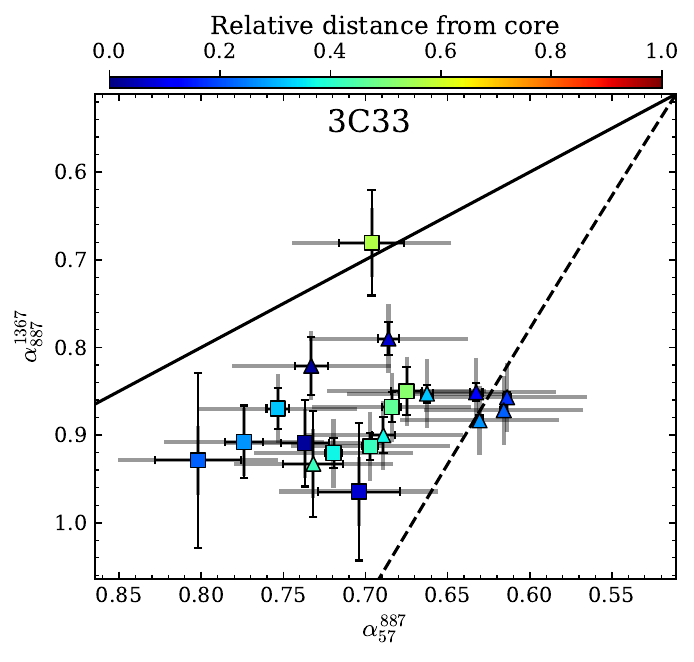}
\includegraphics[width=0.27\linewidth, trim={0.cm 0.cm 0.cm 0.cm},clip]{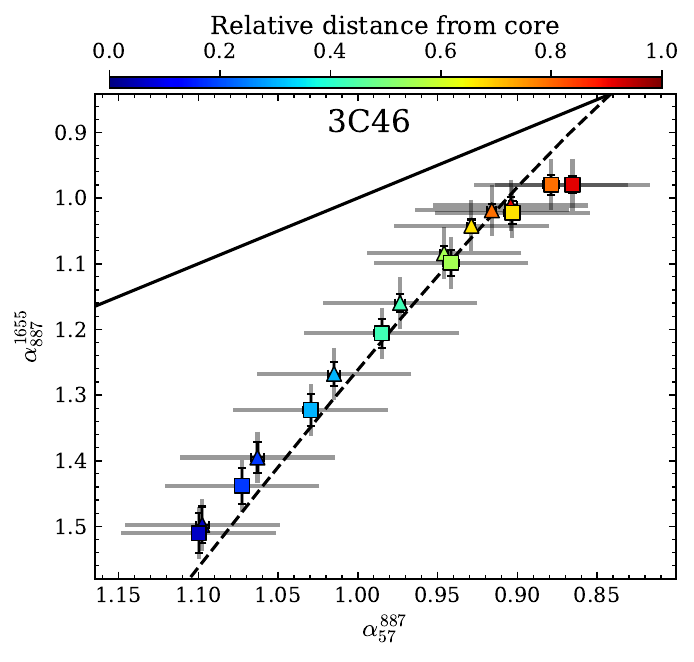}
\includegraphics[width=0.27\linewidth, trim={0.cm 0.cm 0.cm 0.cm},clip]{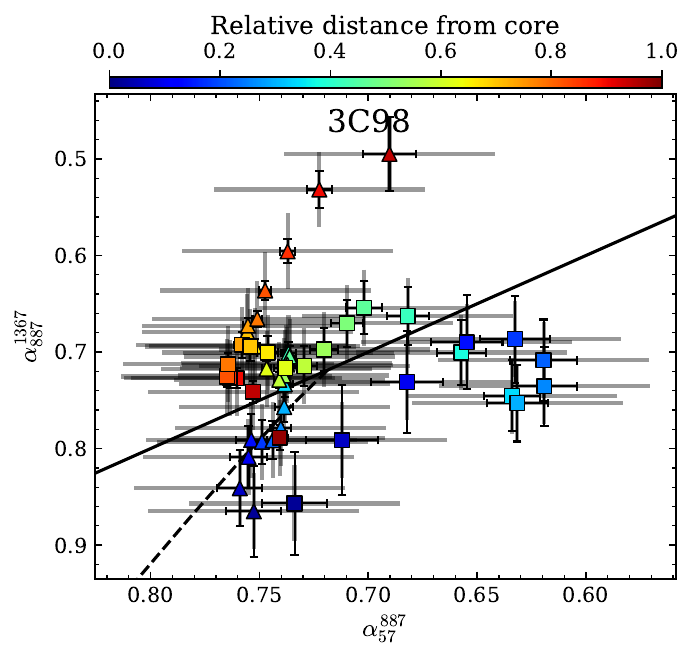}
\includegraphics[width=0.27\linewidth, trim={0.cm 0.cm 0.cm 0.cm},clip]{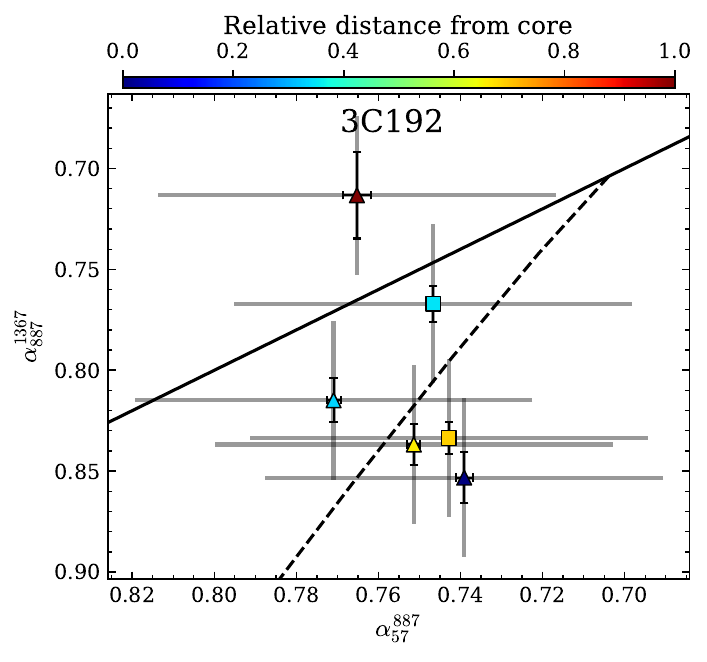}
\includegraphics[width=0.27\linewidth, trim={0.cm 0.cm 0.cm 0.cm},clip]{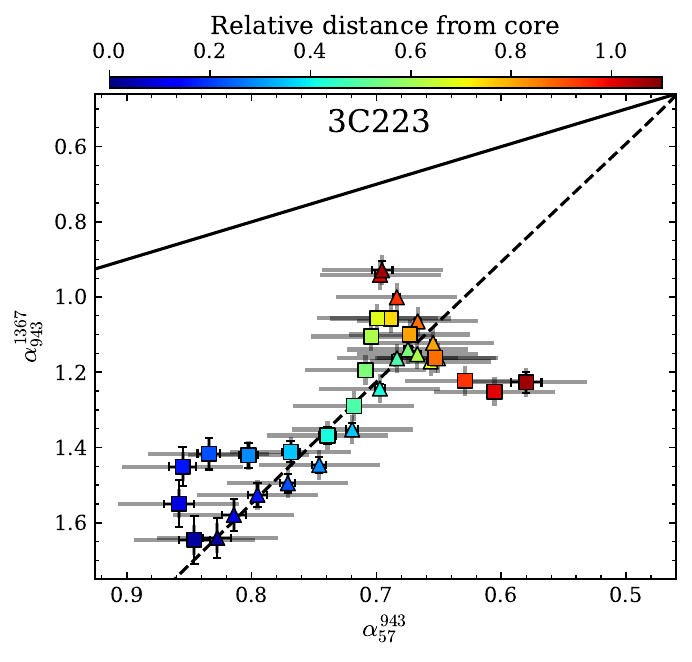}
\includegraphics[width=0.27\linewidth, trim={0.cm 0.cm 0.cm 0.cm},clip]{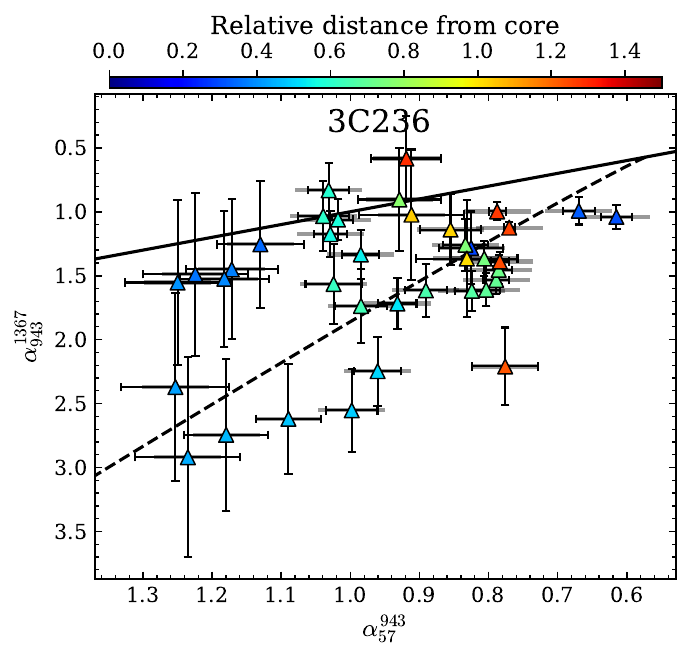}
\includegraphics[width=0.27\linewidth, trim={0.cm 0.cm 0.cm 0.cm},clip]{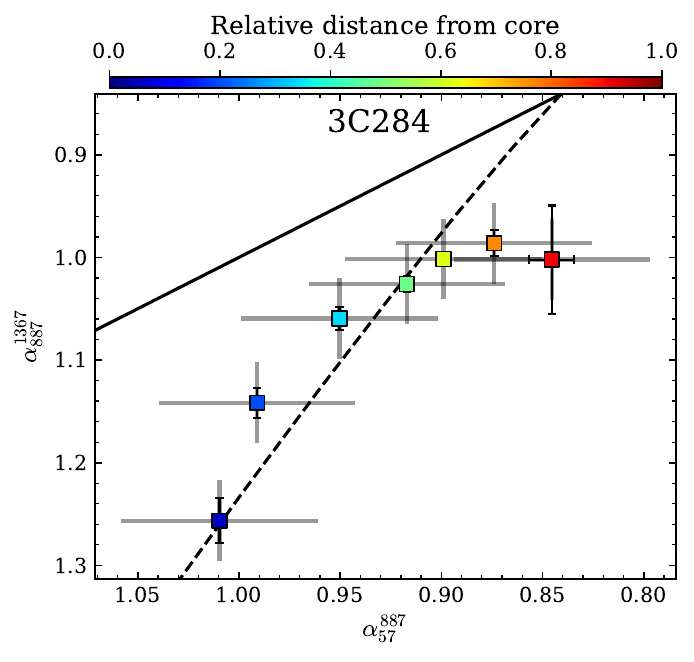}
\includegraphics[width=0.27\linewidth, trim={0.cm 0.cm 0.cm 0.cm},clip]{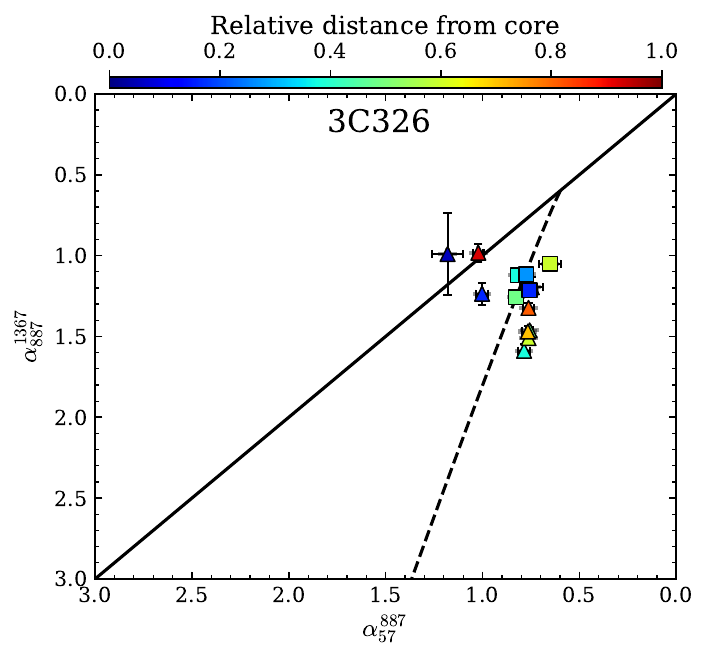}
\includegraphics[width=0.27\linewidth, trim={0.cm 0.cm 0.cm 0.cm},clip]{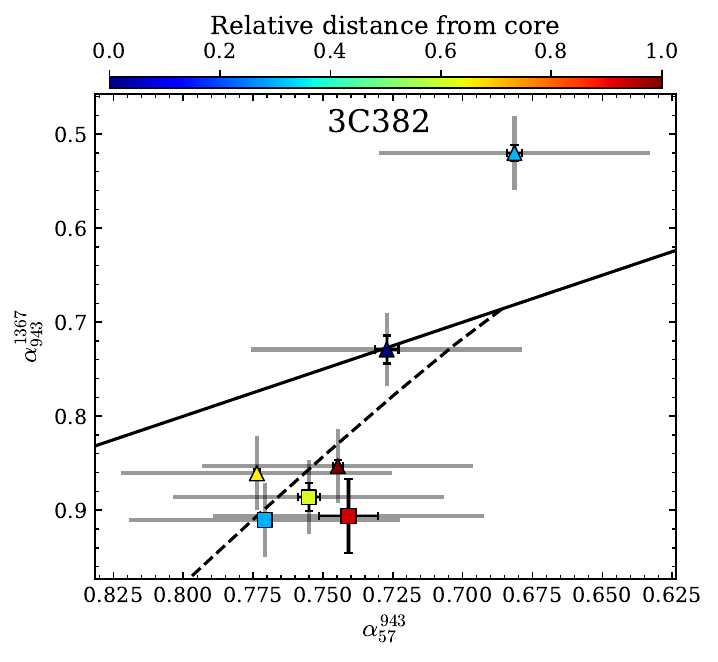}
\includegraphics[width=0.27\linewidth, trim={0.cm 0.cm 0.cm 0.cm},clip]{Assets/color_color/color_color_3c442a___0,2_,_2,3__.pdf}
\includegraphics[width=0.27\linewidth, trim={0.cm 0.cm 0.cm 0.cm},clip]{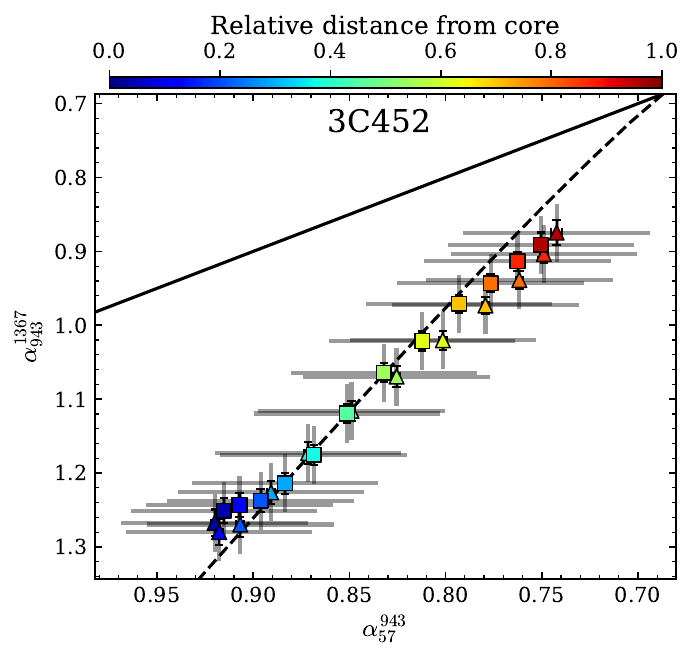}
\includegraphics[width=0.27\linewidth, trim={0.cm 0.cm 0.cm 0.cm},clip]{Assets/color_color/color_color_3c457___0,2_,_2,3__.pdf}
\includegraphics[width=0.27\linewidth, trim={0.cm 0.cm 0.cm 0.cm},clip]{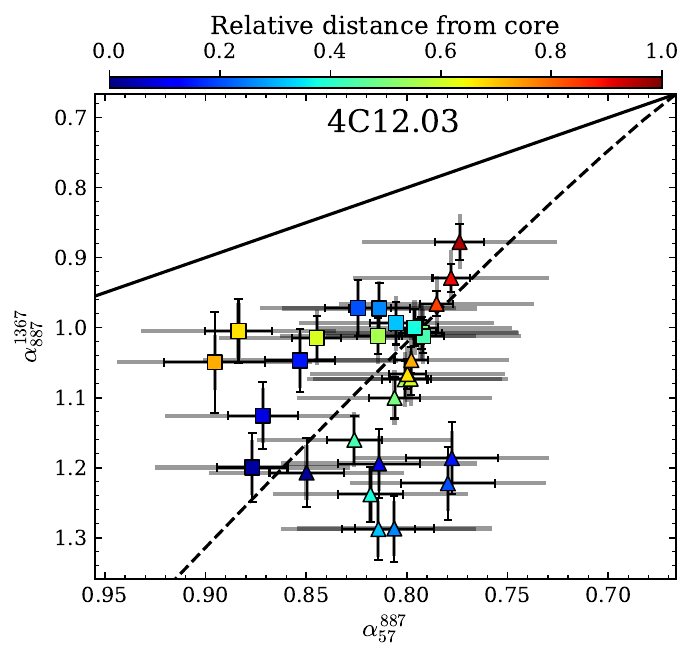}
\caption{Same as Figure~\ref{fig:colour-colour-all-fri}, but for all FR~II in the sample.}
\label{fig:colour-colour-all-frii}
\end{figure}
\end{appendix}

\end{document}